\documentclass[twocolumn]{aastex62}

\usepackage{amsmath}
\usepackage{color}
\usepackage{multirow}
\usepackage{subfigure}
\usepackage{lineno}
\usepackage{graphicx}
\usepackage[toc,page]{appendix}
\usepackage{chemformula}

\begin{document}

\title{Detection of an Atmosphere on a Rocky Exoplanet}

\author{Mark R. Swain}
\affil{Jet Propulsion Laboratory, California Institute of Technology, 4800 Oak Grove Drive, Pasadena, California 91109, USA}
 
\author{Raissa Estrela}
\affil{Jet Propulsion Laboratory, California Institute of Technology, 4800 Oak Grove Drive, Pasadena, California 91109, USA}
\affil{Center for Radio Astronomy and Astrophysics Mackenzie (CRAAM), Mackenzie Presbyterian University, Rua da Consolacao, 896, Sao Paulo, Brazil}
 
\author{Gael M. Roudier}
\affil{Jet Propulsion Laboratory, California Institute of Technology, 4800 Oak Grove Drive, Pasadena, California 91109, USA} 
 
\author{Christophe Sotin}
\affil{Jet Propulsion Laboratory, California Institute of Technology, 4800 Oak Grove Drive, Pasadena, California 91109, USA}

\author{Paul Rimmer}
\affil{Department of Earth Sciences, University of Cambridge, Downing St, Cambridge CB2 3EQ, UK}
\affil{Cavendish Laboratory, University of Cambridge, JJ Thomson Ave, Cambridge CB3 0HE, UK } 
\affil{MRC Laboratory of Molecular Biology, Francis Crick Ave, Cambridge CB2 0QH, UK}

\author{Adriana Valio}
\affil{Center for Radio Astronomy and Astrophysics Mackenzie (CRAAM), Mackenzie Presbyterian University, Rua da Consolacao, 896, Sao Paulo, Brazil}

\author{Robert West}
\affil{Jet Propulsion Laboratory, California Institute of Technology, 4800 Oak Grove Drive, Pasadena, California 91109, USA}

\author{Kyle Pearson}
\affil{Jet Propulsion Laboratory, California Institute of Technology, 4800 Oak Grove Drive, Pasadena, California 91109, USA}

\author{Noah Huber-Feely}
\affil{Columbia University, New York, New York, 10027, USA}
 
\author{Robert T. Zellem}
\affil{Jet Propulsion Laboratory, California Institute of Technology, 4800 Oak Grove Drive, Pasadena, California 91109, USA}

\begin{abstract}

We report the detection of an atmosphere on a rocky exoplanet, GJ 1132 b, which is similar to Earth in terms of size and density. The atmospheric transmission spectrum was detected using Hubble WFC3 measurements and shows spectral signatures of aerosol scattering, HCN, and CH$_{4}$ in a low mean molecular weight atmosphere.  We model the atmospheric loss process and conclude that GJ 1132 b likely lost the original H/He envelope, suggesting that the atmosphere that we detect has been reestablished. We explore the possibility of H$_{2}$ mantle degassing, previously identified as a possibility for this planet by theoretical studies, and find that outgassing from ultrareduced magma could produce the observed atmosphere.  
In this way we use the observed exoplanet transmission spectrum to gain insights into magma composition for a terrestrial planet. The detection of an atmosphere on this rocky planet raises the possibility that the numerous powerfully irradiated Super-Earth planets, believed to be the evaporated cores of Sub-Neptunes, may, under favorable circumstances, host detectable atmospheres.

\end{abstract}


\section{Introduction} 

Stripping of the primordial H/He envelope by photo-evaporation, which is most efficient during the first $\sim$100 My, has been invoked to explain the deficit in the exoplanet occurrence-rate distribution of small planets in short-period orbits \citep{owen2013,fulton2017,owen2017}. The planet occurrence-rate deficit occurs around Rp = 1.5–2.0 R$_{\Earth}$, for planets in short-period orbits that are exposed to relatively high levels of insolation from the parent star. Removal of the envelope leaves behind a bare core of typically $\sim$5 M$_{\Earth}$ \citep{swain2019}, which would require either outgassing or volatile delivery to re-establish an atmosphere. In principle, a sub-Neptune planet could be stripped of the original H/He envelope, become a super-Earth, which then acquires a secondary atmosphere. Due to thermal escape and other loss mechanisms, resupply of H might be needed to maintain a significant H$_{2}$ atmosphere component on terrestrial planets. Recent work has shown that significant amounts of hydrogen, associated with the primordial H/He envelope, can dissolve in the magma ocean of young exoplanets during the envelope-assembly stage and can explain the abundance of sub-Neptune planets \citep{kite2019}. This process provides a mechanism to store primordial H, creating a reservoir of volatile material from which can later be released to replenish atmospheric losses \citep{chachan2018,kite2020}. This potential H supply mechanism is observationally significant because it could lower the mean molecular weight of the planetary atmosphere and thus enhance the observability of the transmission spectrum. The ideal planet to search for signatures of a secondary H$_{2}$ atmosphere would be a strongly irradiated Super Earth that has been observed with the Hubble Telescope, currently the premier facility for characterizing exoplanet atmospheres via the transit spectroscopy method. GJ 1132 b meets these criteria. 

GJ 1132 b has a mass of 1.66 $\pm$ 0.23 M$_{\Earth}$ and a radius of 1.16 $\pm$ 0.11 R$_{\Earth}$, resulting in a mean density of 6.3 $\pm$ 1.3 g cm$^{-3}$, establishing it as a terrestrial planet \citep{berta-thompson2015,bonfils2018}. It orbits a cool M4.5 dwarf star with a period of 1.6 days. GJ 1132 b has an estimated equilibrium temperature \citep{bonfils2018} of 529 $\pm$ 9 K and receives an insolation of 19 times that of Earth, placing it in a population of high-insolation terrestrial planets that, as a group, have been identified as likely having lost the primordial H/He envelope \citep{owen2017,swain2019}. Data from 5 visits of Hubble WFC3-G141 infrared grism transit observations of GJ 1132 b (PI Berta-Thompson, PID 14758) were downloaded from the MAST archive and, using the EXCALIBUR (EXoplanet CALIbration and Bayesian Unified Retrieval) science data pipeline, analyzed these data to produce a near-infrared transmission spectrum of the planet’s atmosphere.  

\section{Methods}

Here we discuss our methods for the data reduction including the data calibration, light curve modeling, and atmospheric retrieval steps. We also discuss our spectral validation methods used to ensure the veracity of the results. 

\subsection{Data Calibration}

The Hubble WFC3-G141 data set used to obtain the infrared transmission spectrum consisted of 5 Hubble visits, of 4 orbits each, observed on the following dates in 2017: 04/18, 09/23, 11/19, 11/23, 11/26. These data were obtained in the bi-directional scan observing mode and provided excellent coverage of ingress, egress, the transit, as well as out-of transit-baseline. 
The data calibration steps needed to obtain the transmission spectrum are performed in the EXCALIBUR science data pipeline. The implementation strategy for EXCALIBUR is to use established data reduction methods where feasible, building upon the work by practitioners such as \cite{deming2013,kreidberg2014,tsiaras2016,tsiaras2019} and others.  There is considerable community experience with Hubble WFC3 data reduction, and we make use of this when ever possible. With respect to previous work, one difference in the EXCALIBUR approach is the capability to operate at the pixel-based spectral resolution of the WFC3 instrument. This capability exploits the similarity between the WFC3-IR plate scale, 0.13$^{\arcsec}$, and point spread function full width half max, 0.136$^{\arcsec}$ at 1.2459 $\mu$m, \citep{windhorst2011}. Independent of the level of spectral smoothing selected for the final science interpretation, in principle, a WFC3-IR pixel-based spectrum allows for improved outlier rejection, by not including highly discrepant pixel-based transit depth values that could otherwise bias the transit depth estimate based on averaging during the spectral extraction.

\begin{figure}[htp!]
\centering
\includegraphics[width=1\columnwidth]{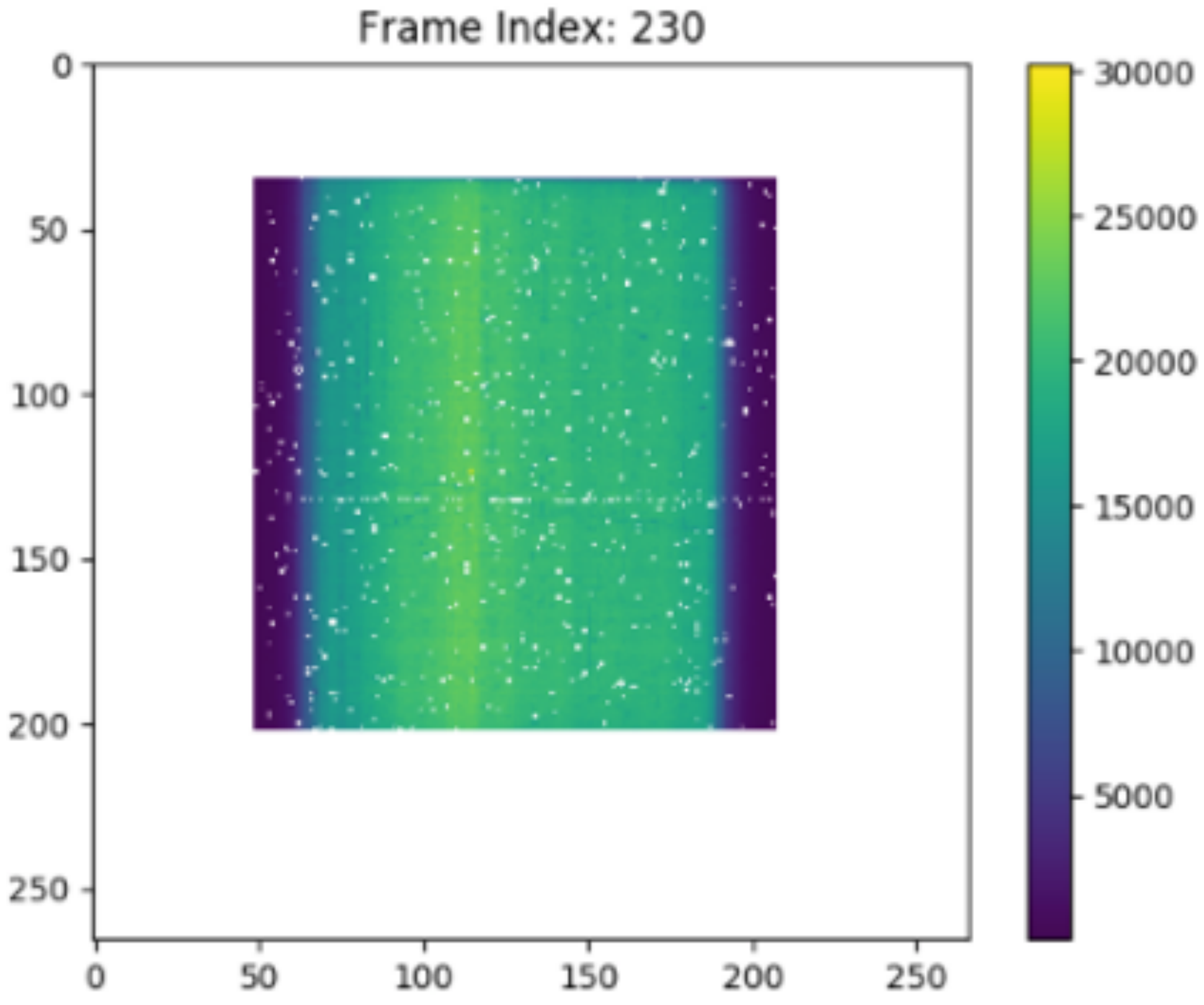}
\caption{An example of the spatially scanned spectrum extracted from a single 103 second WFC3-G141 exposure. The horizontal and vertical units are pixels and the color scale shows detector counts.}
\label{fig:scan} 
\end{figure}

The starting point for our Hubble WFC3 data reduction is the MAST archive .ima data product, which includes calibrations for standard near-infrared detector properties and cosmic ray removal \citep{dressel2017}. We extract the spatially scanned signal following a previously reported method \citep{deming2013} based on using the difference images of successive nondestructive detector reads, and mask the science target and background object pixels. The science target mask position is optimized for each nondestructive read, also termed sampling up the ramp, in an exposure. The median of unmasked pixels is used to subtract residual background in each sample-up-the-ramp difference image. The background-subtracted, sample-up-the-ramp, difference images are then added together to make the spatial scanned, extracted, science target spectrum (see Figure ~\ref{fig:scan}) corresponding to science light collected during the detector integration time. We assemble the data from each orbit into a series of visits. We use the assembled data to define the time of observations, $t$, with respect to three different types of events, $t_{1}=t - T_{0}$, $t_{2}= t-t_{\mu}$, $t_{3} = t - t_{i}$, where $T_{0}$ is the time of mid transit, $t_{\mu}$ is the mean of the time stamps for the individual observation times in an orbit, and $t_{i}$ is the time of the initial observation in a given orbit. The data analyzed here were acquired in dual scan mode where the odd and even detector integrations are scanned in opposite directions. Because instrument systematics depend on the scan direction, we find it convenient to treat the odd and even scans as separate visits. Thus we we define a visit as consisting of a series of Hubble orbits spanning the transit event, during which time a sequence of detector integrations, all spatially scanned in the same scan direction, are acquired. When building the visit time series, we follow the practice of neglecting the first orbit to allow for spacecraft settling \citep{deming2013,kreidberg2014}.

During the detector integration time, 103.128633 seconds for these observations, the spectrum is scanned in a direction that is orthogonal to the dispersion axis; however, scan rate errors, scan drifts in the dispersion direction, and an optical distortion cause the scan area to depart slightly from a uniformly illuminated rectangle \citep{deming2013,tsiaras2016}. Scan rate errors cause non-uniform illumination along the scan direction, while scan drifts in the dispersion direction causes shifts in the spectrum position as the spectrum is scanned along the detector \citep{kreidberg2014}. Additionally, position-dependent optical distortion stretches the spectrum along the dispersion axis \citep{tsiaras2016}.

To correct scan rate errors in each detector integration, we construct a scan-rate template by computing the median value of each row in Figure ~\ref{fig:scan} and then divide each column in the integration by the scan-rate template. We then construct a spectral template by computing the median of each column in Figure ~\ref{fig:scan}. We correct scan-induced position errors in the dispersion direction and optical distortion by fitting for spectral shift and stretch terms using the spectral template. The corrected spatial scan image is now collapsed along the scan axis, by taking the median value of each colunm, to extract the observed spectrum. Rather than taking a sum, we use median values in the template and spectral extraction operations to reduce sensitivity to mask positioning. The extracted spectrum for each integration, corresponding to a single spatial scan measurement, is corrected for the instrument passband filter and wavelength calibrated using the instrument wavelength solution following previously published methods \citep{deming2013}. These spatial scan calibration steps are repeated for the sequence of spatial scans, resulting in a spectral time series. Because the HST scan starting position of each spectral scan image is not exactly repeatable in terms of spectral placement on the detector pixel grid, each wavelength-calibrated spectrum associated with each spatial scan in the time series can have a slightly different wavelength grid. We group the spectral points in bins 4.65 nm wide, corresponding to the WFC3 G141 dispersion per pixel \citep{dressel2017}. 

At this stage of the calibration, changes in flux within a given 4.65 nm spectral channel, between different scans, have components due to both spectral position shifts on the detector and the WFC3-IR detector ramp effect; these two effects must be disentangled to enable interpolation of the spectra from individual scans onto a common wavelength grid for the visit. To estimate the bias due to wavelength-independent, time-correlated effects we model them as:
\begin{align}
S(t) = 1-e^{- \left( \frac{t_{3} + \alpha}{\beta} \right)}
\label{eq:step1}
\end{align}
where $S(t)$ is the normalized broadband, sometimes termed ``whitelight", light curve and $\alpha$ and $\beta$ are parameters with format $10^{x}$.  We apply the model from Equation 1 to the pretransit and post-transit orbits and solve for coefficients $\alpha$ and $\beta$ using a Levenberg–Marquardt method. Thus equipped, we apply the Equation~\ref{eq:step1} model parameters to the entire visit, and then select a uniform wavelength grid that is centered in the data for the visit and interpolate the spectra onto this grid with a second degree polynomial. At this point in the calibration we now have a time series of extracted, calibrated spectra on a common wavelength grid. However, before we can start to fit for a transit depth, we need to produce a model for stellar limb darkening.

\subsection{Light Curve Modeling}
\begin{figure}[htp!]
\centering
\includegraphics[width=1\columnwidth]{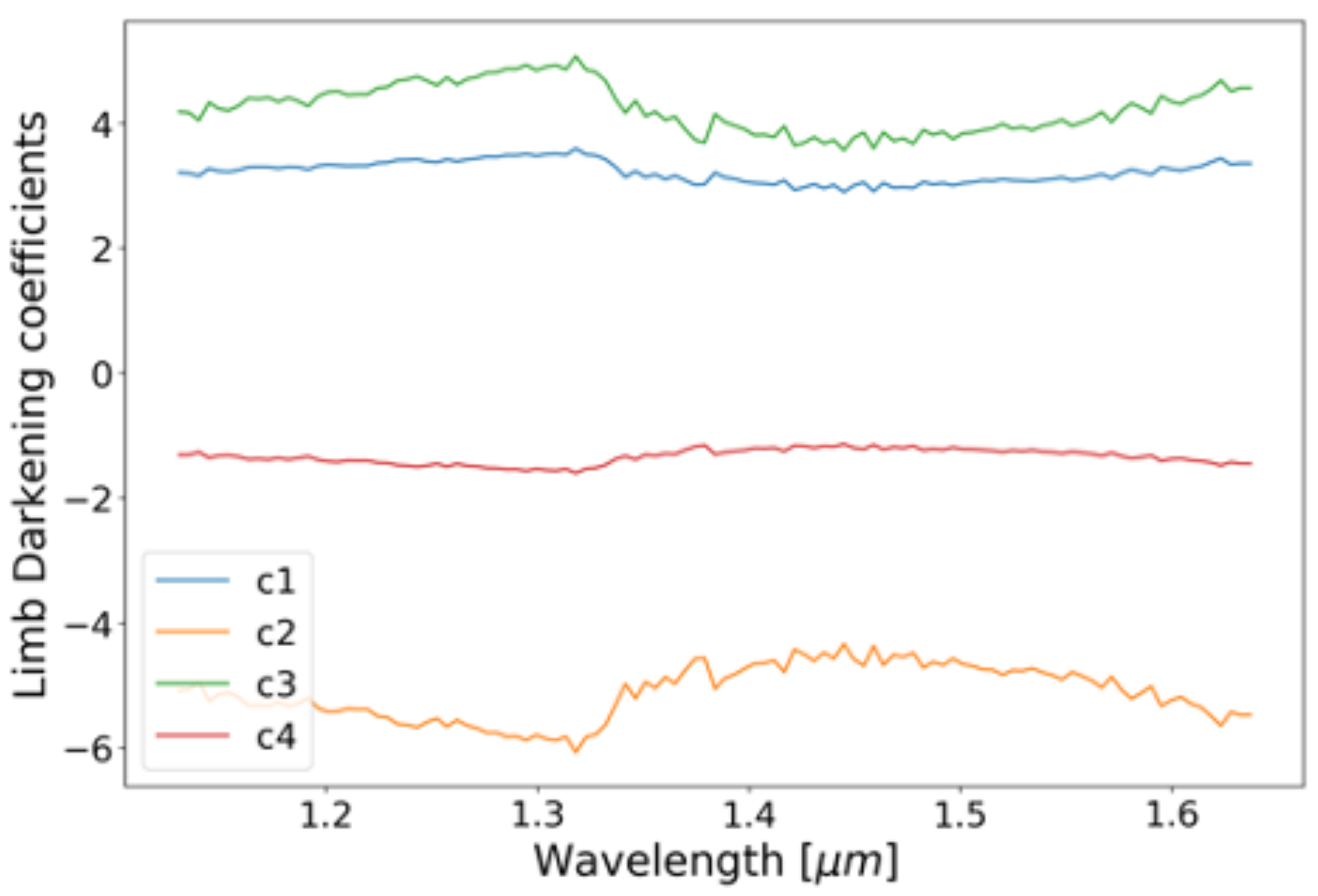}
\caption{The non-linear limb darkening solutions used for modeling the transit light curve, computed at the pixel-based spectral resolution. The coefficient c$_{4}$ is constrained so that the sum of coefficient values is one.}
\label{fig:limbdark}
\end{figure}

The shape of the transit light curve depends on stellar limb darkening. We model this effect using a four-parameter non-linear limb-darkening law \citep{claret2000}: 
\begin{align}
\frac{I(\mu)}{I} = 1 - c_{1} ( 1 - \mu^{\frac{1}{2}}) - c_{2} ( 1 - \mu) - \\ \nonumber
c_{3} (1-\mu^{\frac{3}{2}}) - c_{4}(1-\mu^{2})
\end{align}
where the $c_{n}$ are the limb darkening coefficients and $\mu$ is the cosine of the emission angle. Using the LDTK package \citep{parviainen2015}, we determine the values of $c_{n}$ and their uncertainties using Phoenix grid models \citep{husser2013}, spline interpolated to the stellar temperature, metallicity, and surface gravity of GJ 1132 b. We compute limb darkening coefficients at the pixel-based spectral resolution of the data (see Figure ~\ref{fig:limbdark}). The coefficient $c_{4}$ is treated as a condition on the coefficients' sum so that $I(\mu\!\!=\!\!0)\!=\!0$, corresponding to full extinction at the end of the limb, which ensures proper normalization of the stellar disk (see Figure ~\ref{fig:limbdark}). The limb darkening parameters are calculated at the pixel-based spectral resolution of the WFC3-G141 data. We implement the limb darkening calculation in our own transit light curve, LC, modeling code validated by comparing to results from \cite{mandel2002}. The LC parameterizes the transit in terms of the square of the radius of the planet divided by the radius of the star, $(R_{p}/R_{s})^{2}$, and uses the exoplanet system parameters together with the limb-darkening relationship. In addition to modeling the stellar contribution to the transit light curve shape, we also need to accurately determine the system parameters.

We determined system parameters consistent with the Hubble/WFC3 observations by constructing a broadband (1.15–1.65 $\mu$m) light curve consisting of all the visits, incorporating an outlier rejection algorithm acting on the time-series spectra that flags wavelength channels of the spectrum that are not stable relative to 3 times the expected shot-noise level in the out-of-transit portion of the signal. We model the normalized broadband transit light curve, $S(t)$, using orbital parameters from the NASA Exoplanet Science Institute (NExScI) database (see Table ~\ref{tab:system}) and incorporate the instrument model, including limb darkening using: 
\begin{align}
S(t) = LC \underbrace{\left[ v_{s} t_{1} + 1 \right]}_\text{visit} \times \\ \nonumber
\underbrace{\left[ \left(1-e^{-\frac{t_{3} + \alpha}{\beta}} \right) \left[ o_{s} t_{2} + o_{i} \right] \right]}_\text{orbit}
\label{step2}
\end{align}
where $v_{s}, o_{s}, o_{i}$ are slope and intercept parameters for either visit or orbit as indicated above. The parameters $v_{s}, o_{s}, o_{i}$ are constrained to take on the same value for a given visit while the parameters $\alpha, \beta$ have been previously determined. The broadband transit light curve model is used in conjunction with a Metropolis-Hastings sampler from the PyMC3 package \citep{salvatier2016} to retrieve the broadband instrument linear model parameters, the normalized planetary radius, $R_{p}/R_{s}$, an improved values for the inclination, and time of mid transit (see Table ~\ref{tab:system}). These retrieved values can now be applied to the spectral light curve modeling process.

\begin{figure*}[htp!]
\centering
\includegraphics[width=1.7\columnwidth]{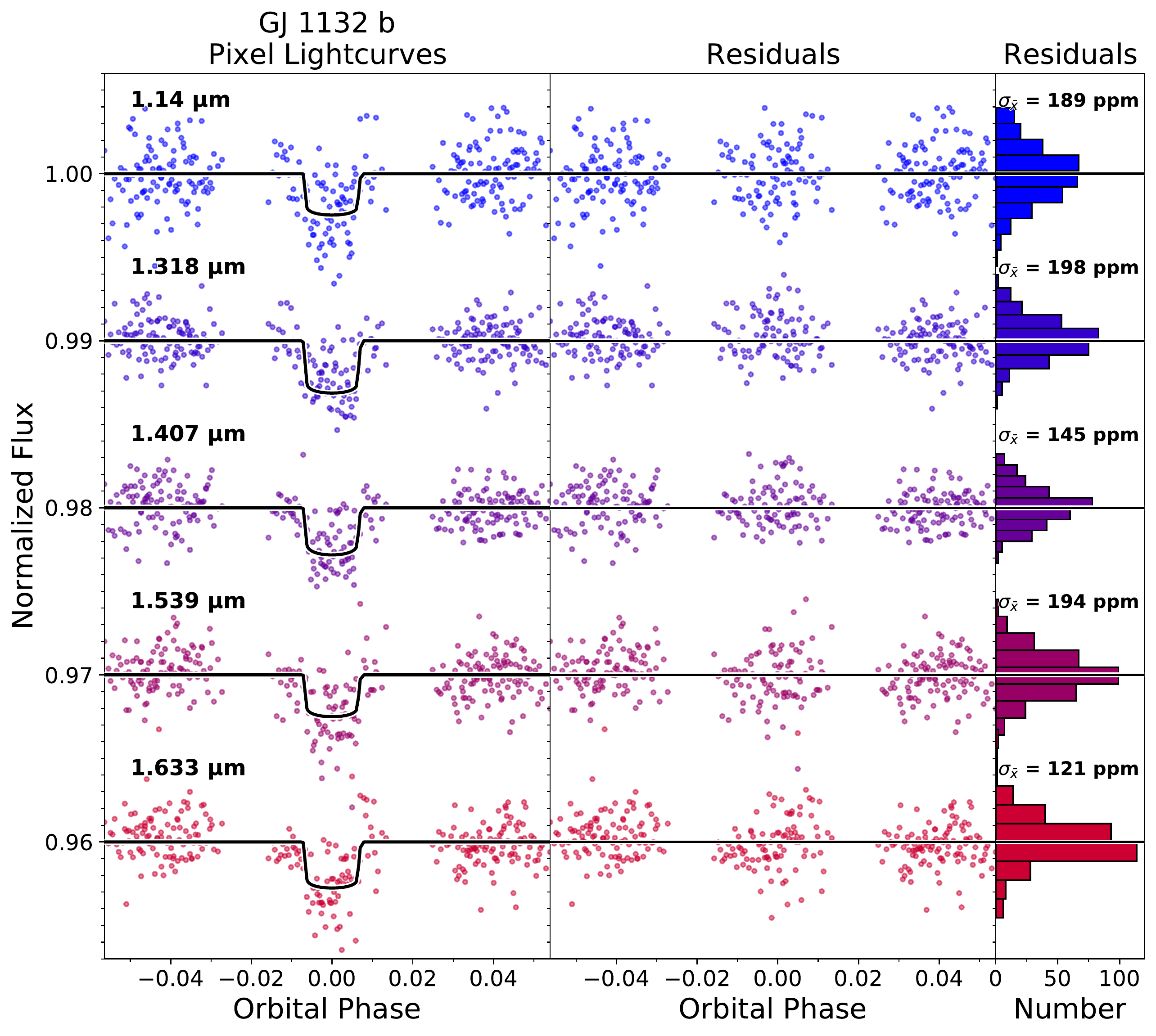}
\caption{Examples of single-pixel transit light curves corresponding to the grey points in Figure ~\ref{fig:spectrum} are shown  (column 1) together with the light-curve model residuals displayed as a time series (column 2) and as histograms (column 3) and standard deviation in the mean for the residuals.  The data are plotted with no temporal averaging.} 
\label{fig:lightcurves}
\end{figure*}

For retrieving the transmission spectrum, we follow a similar approach to modeling the broadband light curve except that we now use the retrieved orbital solution values for the semi major axis, impact parameter, and time of mid transit. We construct spectral light curves using data from all 5 visits and apply the same outlier rejection method as for the broad band light curve. We model the normalized spectral light curves by generalizing Equation ~\ref{step2} for the individual spectral light curves:
\begin{align}
S_{\lambda}(t) = LC_{\lambda} \left[ v_{s,\lambda} t_{1} + 1 \right] \times \\ \nonumber
\left[ \left(1-e^{-\frac{t_{3} + \alpha}{\beta}} \right) \left[ o_{s,\lambda} t_{2} + o_{i,\lambda} \right] \right]
\label{step3}
\end{align}
where the subscript $\lambda$ denotes the wavelength dependence of the astrophysical light curve and the instrument model parameter coefficients. As in the orbital solution step, we use the PyMC3 package \citep{salvatier2016} to retrieve the wavelength-dependent instrument model parameters as well as the normalized planetary radius $R_{p}/R_{s}$ on the individual spectral light curves at the pixel-based spectral resolution of the measurements. Our normally distributed prior for the planetary radius is centered on the broadband recovered planetary radius, with a full width half max of 5 Hs, appropriate for the G141 band at this spectral resolution. At spectrum level we flag spectral channels that have a spectral transit depth of $\pm$ 5 atmospheric scale heights, $H_{s}$, the broadband transit depth. In Figure ~\ref{fig:lightcurves} we show examples of individual pixel-based spectral light curves.

\begin{table*}[htp!]
\centering
\begin{tabular}{|l|l|l|l|l|l|l|}
\hline
\textbf{parameter} &  \textbf{value} & \textbf{Upper limit} & \textbf{Lower limit} & \textbf{units} & \textbf{type} & \textbf{reference}\\
\hline
inclination & 86.58  &  0.63  & -0.63 & deg & prior & \citep{southworth2017} \\ 
\hline
time of mid-transit & 2457184.5576  & 1/2 duration  & 1/2 duration & MJD & prior & \citep{southworth2017} \\ 
\hline
period & 1.628931 & NA & NA & MJD & fixed & \citep{bonfils2018} \\
\hline
semi-major axis & 0.0153 & NA & BA & au & fixed & \citep{bonfils2018} \\
\hline
inclination & 87.3577 & 0.0430105 & 0.044457 & deg & retrieved & this work\\
\hline
time of mid-transit & 2457184.0562274 & 0.0001602 & 0.0001602 & MJD & retrieved & this work\\
\hline
\end{tabular}
\caption{The orbital parameters and values used for the light curve fit. Those labeled as retrieved are estimated as part of our analysis using the broad band light curve from the multi-visit WFC3-G141 transit observations of GJ 1132 b.}
\label{tab:system}
\end{table*}

\subsection{Atmospheric Retrieval}

The retrieval of atmospheric model parameters from the WFC3 G141 observations, is done with Cerberus, as a line by line, spherical shell geometry radiative transfer code. Cerberus implements the formalism of \cite{tinetti2012} in a Python software that couples forward modeling with parameter retrieval using a Metropolis Hastings Markov Chain Monte Carlo (MCMC) sampler. Cerberus uses the MCMC package from PyMC3 \citep{salvatier2016}.

Following \cite{benneke2012}, we define the planet-to-star radius, assume a solid radius at 10 bars \citep{griffith2014}, and extend the atmosphere through 20 scale heights, using a hundred isothermal pressure layers in hydrostatic equilibrium, evenly distributed in log(pressure). Gaseous sources are modeled based on opacities from EXOMol \citep{tennyson2016} (H$_{2}$O, CH$_{4}$, HCN, H$_{2}$CO, CO, CO$_{2}$), HITEMP \citep{rothman2010} (C$_{2}$H$_{2}$, N$_{2}$, N$_{2}$O, O$_{2}$, O$_{3}$, TiO) and HITRAN \citep{richard2012} (H$_{2}–$H$_{2}$, H$_{2}–$He, H$_{2}$–H, He–H), and we assume a constant vertical mixing ratio (VMR) for gaseous species. The effect of pressure on line shifting and broadening is modeled following a previously described method \citep{rothman98}, and the cross-section temperature dependence is represented using the HITRAN/HITEMP internal partition function \citep{fischer2003}. In the retrieval process, each gas specie is described by a mixing ratio and distributed homogeneously throughout all atmospheric layers, with a mixing ratio value that is either computed from the usual parameterization of thermal equilibrium chemistry (Metallicity, C/O, N/O) or treated as a free parameter. The mixing ratios for the gas species tracked are summed and the balance of the atmosphere then filled with H/H$_{2}$ in solar relative ratio. Rayleigh scattering by H$_{2}$ is modeled using a scaled power law \citep{naus2000}, with an index of -4, to account for the wavelength-dependent cross-section.

We also model the contributions of clouds and aerosols in the retrieval. We model clouds as fully opaque, obscuring all pressure layers below the Cloud Top Pressure (CTP) level, set up as a free parameter, expressed in log($p$) where $p$ is pressure. We model scattering by parameterizing a vertical haze profile using an aggregate of Jupiter's haze profiles \citep{zhang_x2013} constructed by taking the median of the haze latitude number density. We allow the maximum density location of the haze profile, HLoc, which is a function of log($p$), to linearly translate vertically in the atmosphere. We then introduce a vertical scaling factor factor for the density profile, HThick, which acts to compress or stretch the vertical extent of the initial profile. For reference, a HThick value of 10 results in an approximately uniform density vertical distribution over the range of pressures included in the model. We use Jupiter averaged haze cross section as a reference \citep{west2004}, then scale the haze optical depth for each pressure layer (product of the cross section, density profile and optical path length) with the free parameter HScale, which takes the form log$_{10}(x)$. HScale acts as an adjusting factor to the Jupiter haze cross section and/or density profile. HScale above 10 times the haze optical depth of Jupiter (initial vertical coverage and location) typically renders an opaque atmosphere for the transit viewing geometry.

For modeling the atmosphere of GJ 1132 b, we used a free retrieval approach where the vertical mixing ratio for the main gaseous opacity sources is allowed to take on values that depart from strict Thermal Equilibrium Chemistry. Based on forward modeling at the WFC3 spectral resolution for the range of plausible molecular opacity sources detectable in the G141 wavelength range, we exclude from the retrieval channels with a transit depth ratio departing more than $\pm$ 4 scale heights away from the broadband transit or more than 2.5 scale heights away from a nearest neighbor. After an initial search for opacity sources, we concentrate on the retrieval of molecular opacity sources that have an observational signature in the measured spectrum. The results for our spectral retrieval are presented, together with the observed spectrum, in the following section. In addition to the retrieval activity, where gaseous species are modeled with a constant vertical mixing ratio, we also undertook forward modeling based on physically motivated vertical mixing ratio profiles. In this forward modeling use case, we adopted the output of thermo/photochemical modeling to specify the vertical mixing ratios of various gaseous species. The thermo/photochemical modeling used to generate the molecular species and associated vertical mixing ratios, along with the science motivation, is discussed  in Section 5.2.

\subsection{Spectral Validation}

\begin{figure}[htp!]
\centering
\includegraphics[angle=-90,width=1\columnwidth]{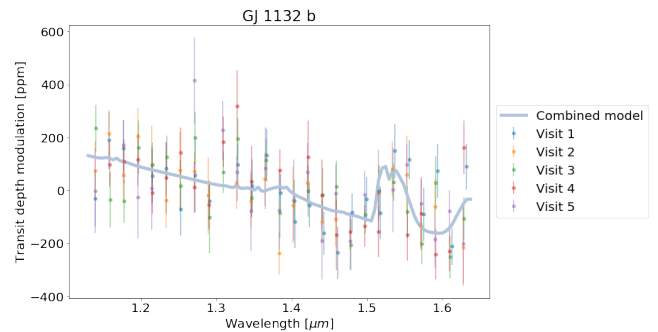}
\caption{The transmission spectrum of GJ 1132 b is repeatable for each of the five visits. Data in individual visits are plotted in different colors and the atmospheric model retrieved from these observations is overplotted in blue.}
\label{fig:5visit}
\end{figure}

Transit observations with multiple visits provide the opportunity for spectral validation by confirming that the  individual spectra are mutually consistent. The WFC3 observations reported on here consist of five separate transit measurements of  GJ 1132 b. We processed the transit spectrta for the five visits individually, including fully independent normalization, visit systematic model parameters, and transit light curve depth determinations. We find the individual visit spectra are in agreement and repeatable as shown in Figure ~\ref{fig:5visit}.

\begin{figure}[htp!]
\centering
\includegraphics[width=1\columnwidth]{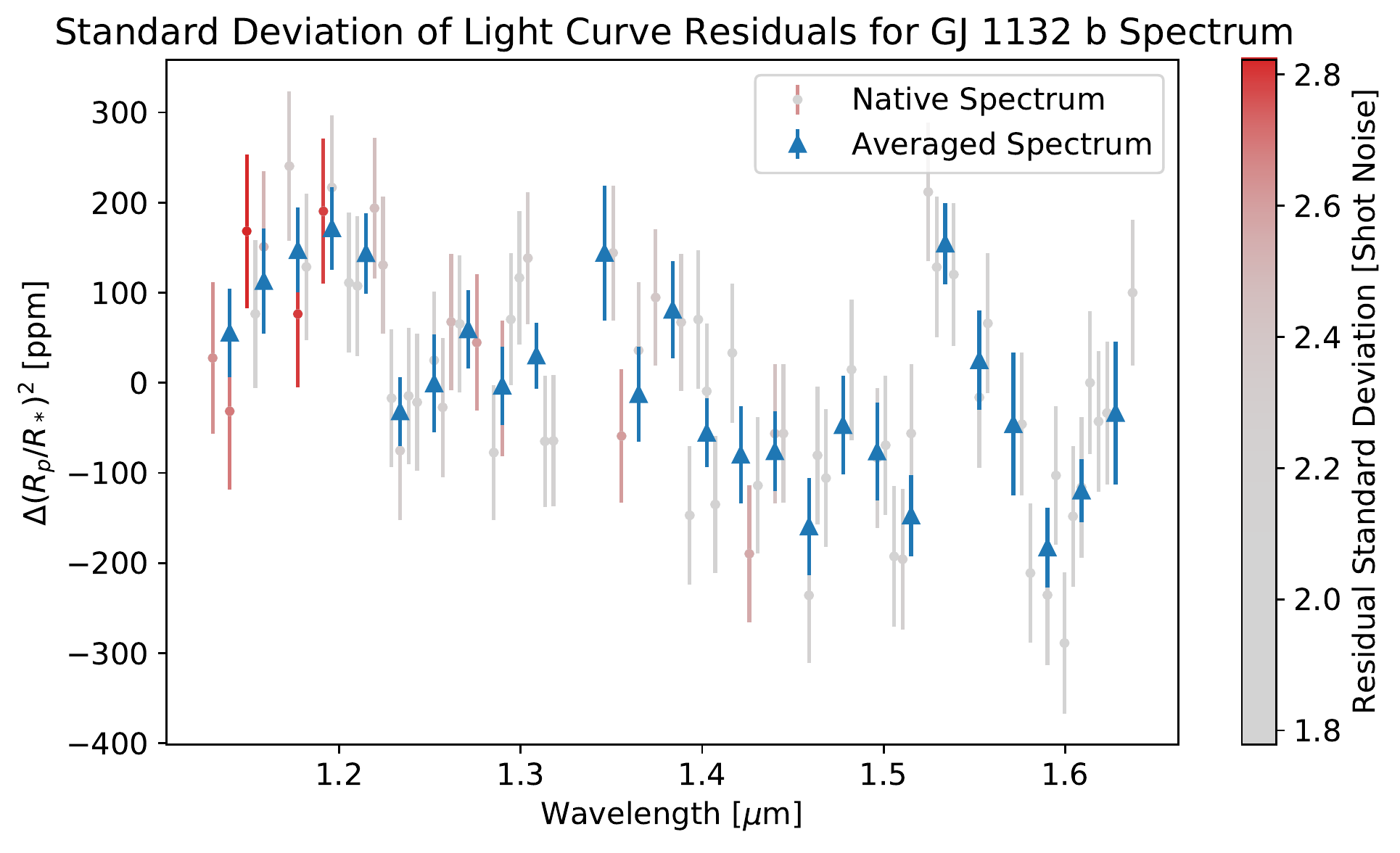}
\caption{The overall quality of the transmission spectrum of GJ 1132 b can be visualized by plotting the spectral channels and displaying the marginalized light curve residual standard deviation in units of shot noise as a color scale. This analysis shows that the majority of the individual pixel-based spectral channels have low levels of residuals.}
\label{fig:residuals}
\end{figure}

We also analyze the residuals for the individual channel light curves to test the hypothesis that the detected spectral features are data artifacts. We find that the overall quality of the individual pixel-based spectral channels is very good, typically about two times the shot noise. Additionally, the few spectral channels with slightly elevated residuals are not associated with the main spectral features we analyze (see Figure ~\ref{fig:residuals}). Thus, we conclude that our result for the spectrum of GJ 1132 b is repeatable and is not caused by a discrepant visit or data outliers.

We also explored the hypothesis that star spots could be the origin of the features present in the WFC3 transmission spectra (see Figure~\ref{fig:5visit}). Ground-based monitoring of GJ 1132 shows variability at visible wavelengths that is $\sim \pm 0.07$ magnitudes \citep{newton2018}, indicating that star spots are present. The level of optical variability reported by \cite{newton2018} is similar to the other M dwarfs they studied and is also consistent with the M dwarf activity reported by \cite{ciardi2011}.  The WFC3 light curves show no indications of star spot crossing, but unocculted starspots can influence transmission spectra given a combination of high precision measurements, deep transits, and an active star \citep{zellem2017}. However, as we discuss below, this combination of factors does not apply to the WFC3 observations of GJ 1132 b, and we identify two lines of reasoning that argue against unocculted star spots being the cause of the features observed in the transmission spectrum.  
\begin{itemize}
\item{\bf Variability amplitude:} \cite{zellem2017} has modeled the spectral impact of unocculted star spots on the transmission spectrum of GJ 1214 b that orbits a host star of the same spectral class \citep{berta2012} as GJ 1132 b. The modeling shows that unocculted starspots have a 11 ppm impact on the transmission spectrum of GJ 1214 b. Since GJ 1214 b and GJ 1132 b are the same spectral class, we can scale the modeling results by the ratio of the measured optical variability of the stars. The variability of GJ 1214 has been reported as $\approx 11$mmag \citep{zellem2017} based on measurements by \cite{nascimbeni2015}, and the variability of GJ 1132 is $\approx7$mmag is based on measurements by \cite{newton2018}.  Thus, the impact of unocculted star spots on the transmission spectrum of GJ 1132 b is $\sim7$ ppm, which is far smaller that the GJ 1132 b transmission spectrum uncertainty of even the 5-visit spectrally binned data points that have an average 1 $\sigma$ uncertainty of 56 ppm.
\item {\bf Measurement repeatability:} the repeatability of the WFC3 spectral slope measurement for five visits (occurring between April 18 and November 26 of 2017) indicates unocculted star spots are not impacting the measurement because, as the star rotates, different transit observations sample different parts of the stellar photoshpere. We find no detectable difference in individual transit depths, again indicating that unocculted spots are not impacting the measurements at a significant level. We also note that while there was no explicit discussion of the potential impact of unocculted starspots or stellar variability in the analysis of five transits \citep{diamond-lowe2018} of spectroscopic measurements between 0.7 and 1.0 $\mu$m that achieve a measurement precision of 90 ppm, there is no indication of transit depth variability presented in those results. 
\end{itemize}
Based on the combination of the impacts of unocculted spots being less than the measurement errors and the repeatability of the WFCE spectral measurements, we conclude that stellar variability is not a significant impact on the WFC3 results. Indeed, with an estimated impact on the spectrum of $\sim12$ ppm, even JWST transmission spectra would not be affected by unocculted star spots unless many visits and extensive spectral averaging were combined.

\section{Results}

Here we discuss our result for the transmission spectrum and modeling. We also compare our findings to previous observational results.

\subsection{Transmission Spectrum and Modeling}

\begin{figure*}[htp!]
\centering
\includegraphics[angle=0,width=2\columnwidth]{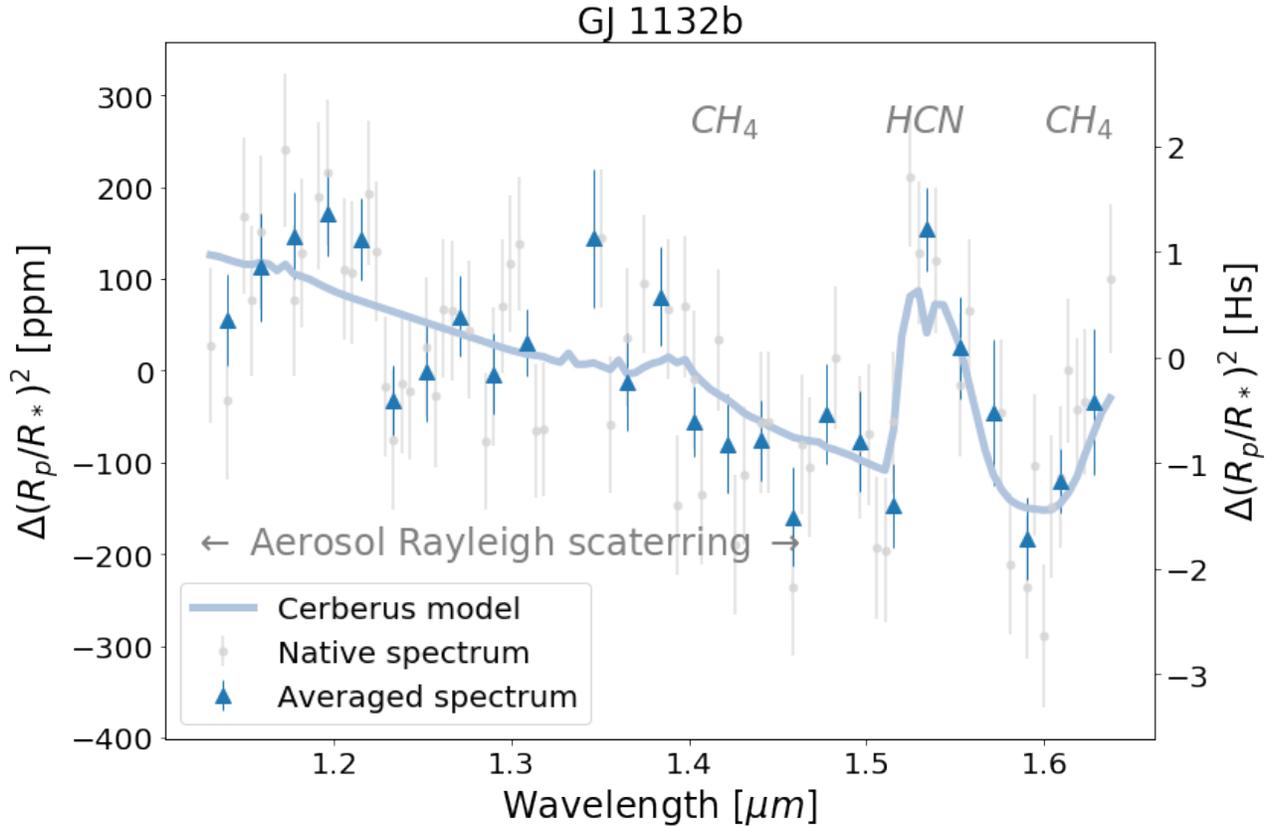}
\caption{The transmission spectrum of GJ 1132 b is consistent with a predominantly H$_{2}$ atmosphere augmented by an aerosol and $\sim$0.5\% each of CH$_{4}$ and HCN, the latter being an indicator of photochemistry.  A few percent of a spectrally inactive species, such as N$_{2}$, could be present, but the total amount of spectral modulation shows the atmosphere remains H$_{2}$-dominated with a low mean molecular weight. This composition is consistent with mantle outgassing. The transmission spectrum data shown here are given in the Appendix.}
\label{fig:spectrum}
\end{figure*}

Over the measurement spectral range, between 1.15 and 1.65 $\mu$m, the near-infrared transmission spectrum (Figure ~\ref{fig:spectrum}) of GJ 1132 b is dominated by three spectral features consisting of a negative slope from 1.15 to 1.6 $\mu$m, a feature suggestive of a molecular band centered at 1.53 $\mu$m, and a positive spectral slope between 1.6 and 1.65 $\mu$m. 

\begin{figure*}[htp!]
\centering
\includegraphics[width=1.1\textwidth]{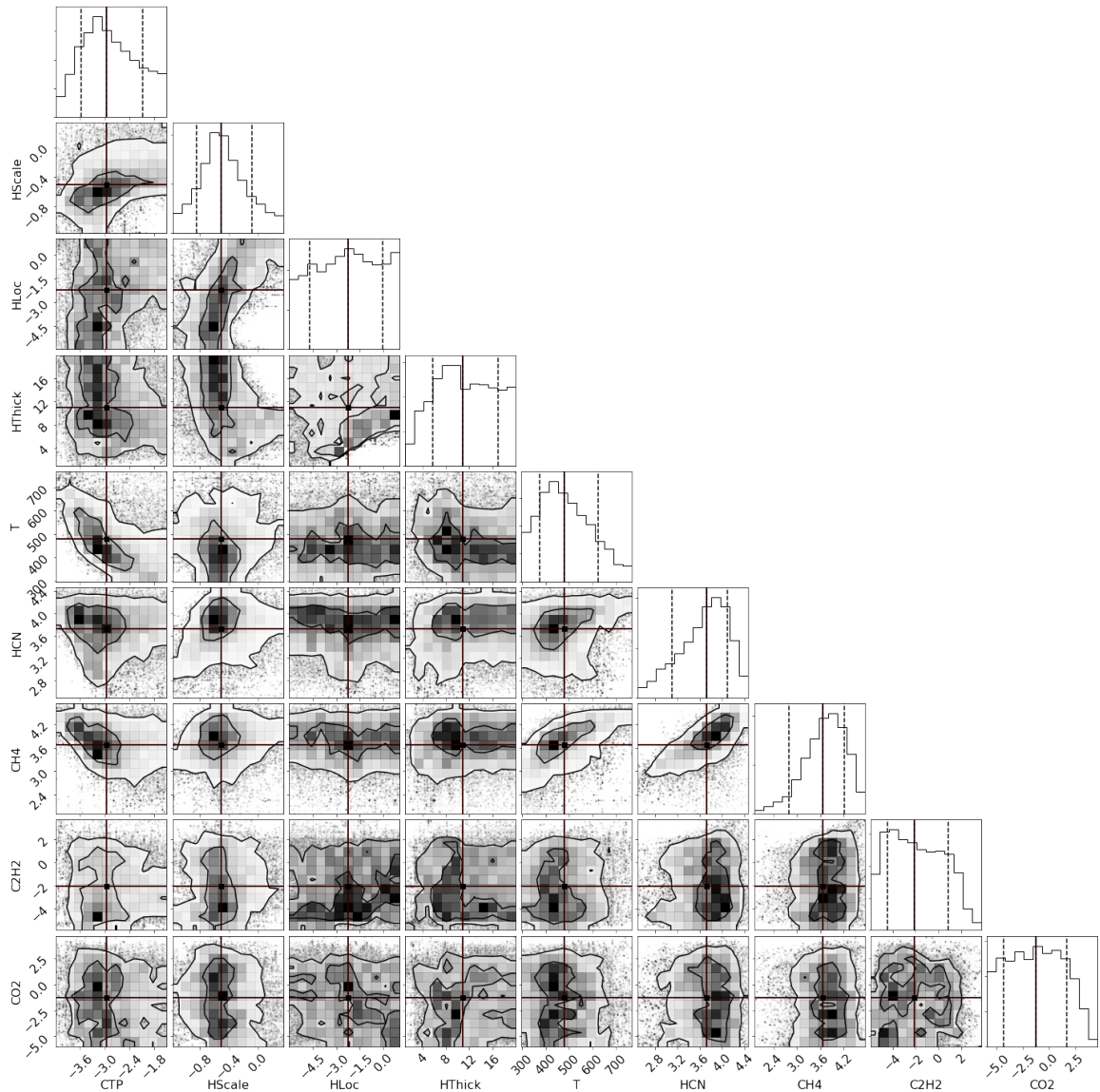}
\caption{The correlation plots and marginalized posterior distributions for retrieved atmospheric model parameters showing detections of CH$_{4}$ and HCN. CTP indicates cloud top pressure. HScale, HLoc, and HThick are all haze parameters corresponding, respectively, to the multiplicative scale factor (log$_{10}$) of Jupiter’s haze-density profile, the maximum density location in long pressure (bars), and a multiplicative factor (log$_{10}$) of the width of the Jupiter haze. T corresponds to the retrieved atmospheric temperature (K), and HCN, CH$_{4}$, C$_{2}$H$_{2}$, and CO$_{2}$ label values correspond to the mixing ratio (log$_{10}$) of each species in units of ppm. The vertical lines correspond to the average (solid) and $\pm$ 1 $\sigma$ confidence interval values (dashed).} 
\label{fig:correlation}
\end{figure*}

We used the Cerberus atmospheric retrieval code, to infer the atmospheric content. Attempting to model the spectrum based on applying thermal equilibrium chemistry to a composition-based scaling metallicity, C/O, and N/O, was not successful. However, we can model the spectrum as a combination of H$_{2}$, HCN, CH$_{4}$, C$_{2}$H$_{2}$, and an aerosol, where each species has its volume mixing ratio as a degree of freedom in the retrieval. Retrieval based on this composition results in a $480_{-105}^{+145}$ K isothermal atmosphere composed of $\sim$ 99$\%$ H$_{2}$, a combination of HCN and CH$_{4}$ totaling $\sim$1$\%$, an aerosol, and a trace amount of C$_{2}$H$_{2}$ with an upper limit of $\sim$1 ppm. The spectrum exhibits a Rayleigh scattering signature in most of the G141 bandpass. We rule out molecular hydrogen as a source of the Rayleigh scattering, due to the forward model providing a poor fit to the observed spectrum in the  H$_{2}$, HCN, CH$_{4}$ and aerosol free atmosphere scenario. The presence of CH$_{4}$ in the atmosphere is inferred from the portion of the spectrum longward of 1.6 $\mu$m. The feature centered around 1.53 $\mu$m is attributed to HCN, based on comparison with C$_{2}$H$_{2}$ as the alternative. The value of atmospheric parameters derived from the spectral retrieval is given in Table ~\ref{tab:retrieval_values}. The retrieval parameter correlation plots (Figure ~\ref{fig:correlation}) show strongly peaked detections for the haze (Hscale), CH$_{4}$, and HCN opacity sources. 

From an observational perspective, the detection of the spectral features associated with aerosols, HCN, and  CH$_{4}$ are robust and is confirmed by measurement repeatability for the individual visits and residual channel variance analysis (see Figure ~\ref{fig:residuals}). Additionally, the presence of H$_{2}$ is consistent with the observations due to the overall spectral modulation; we tested and confirmed this by replacing the H$_{2}$ in our atmospheric model with N$_{2}$, which suppressed the overall spectral modulation and produced a result inconsistent with the observations. Forward modeling implies the observations require an atmospheric molecular weight of $\leq$5. In principal the atmosphere could be He dominated, but previous work \citep{hu2015} demonstrates this scenario is implausible because of the small mass of GJ 1132 b. We conclude the atmosphere is H$_{2}$ dominated, although some He could potentially be present. 

The retrieval of atmospheric model parameters finds well-defined constraints on the vertical mixing ratios for both CH$_{4}$ and HCN (together provide the remaining $\sim$1\%), the presence of an aerosol, and the atmospheric temperature. The presence of haze could be hiding an H$_{2}$O band, and we return to the discussion of determining an upper bound on the H$_{2}$O mixing ratio below. The retrieval includes a cloud component, separate from the aerosol model component, and the retrieval finds a cloud-top pressure at $\sim$1 mbar. Whether this cloud extends beyond the terminator region of the planet’s atmosphere is unconstrained by these observations. Correlation plots and marginalized posteriors for retrieval parameters are shown in Figure ~\ref{fig:correlation}. Our team also analyzed the Spitzer secondary eclipse observations for GJ 1132 b that probe emission from the planet dayside. We find no detectable eclipse in Spitzer IRAC observations, which is consistent with both the equilibrium and retrieval temperature estimates and places an upper limit on the dayside brightness temperature of about 1000 K. In the H$_{2}$ dominated atmosphere scenario used in the retrieval, the atmospheric mean molecular weight is $\sim$2.3 AMU, which implies the spectral modulation corresponds to about 3 atmospheric scale heights (the plausibility for larger values for the atmosphere mean molecular weight is explored in Section 4.2) . The opacity sources in the spectrum, haze + CH$_{4}$ + HCN, coupled with the non-detection of any oxygen-bearing species, suggest enhancement of C, and theoretical studies \citep{hu2014} suggest that the detection of CH$_{4}$ in the atmosphere of GJ 1132 b implies XH $>$ 0.3 and XC/XO $>$ 0.5. The presences of an aerosol may also serve as an indication of enhanced C/O value. Numerous examples of photochemical haze exist in our Solar System, with notable examples including Titan \citep{robinson2014}, Jupiter \citep{zhang_x2013}, and Pluto \citep{zhang_x2017}; in the case of Titan, the haze particles have been simulated in laboratory conditions \citep{gudipati2013}. Laboratory studies of haze formation in hot Jupiter atmospheres show that photochemical haze formation occurs when the C/O $\sim~1$ and is less efficient for C/O $<$ 1 \citep{fleury2019,fleury2020} and performing similar experiments  for temperatures more representative of the conditions in GJ 1132 b is needed. However, measurements with broader wavelength coverage will be needed to observationally establish the atmospheric C/O value for GJ 1132 b.

We considered some possible alternatives to the atmospheric composition identified in the retrieval analysis. The alternative scenarios included a aerosol free atmosphere, an N$_{2}$ rich atmosphere, and an enhanced H$_{2}$O atmosphere. These scenarios were explored using CERBERUS to generate forward models for the hypothesis under test. The models were then compared with the data using $\chi^{2}$ as a figure of merit. Sections 4.2 and 4.3 discuss the alternative atmospheric composition scenarios in more depth as part of a detailed modeling effort to understand the physical origin of the atmosphere detected on GJ 1132 b.

\begin{table*}[htp!]
\centering
\begin{tabular}{|l|c|c|c|c|}
\hline
\textbf{parameter} &  \textbf{label} & \textbf{value} & \textbf{16$^{th}$ percentile} & \textbf{84$^{th}$ percentile} \\
\hline
cloud top pressure & CTP  &  1.1 mbar  & $+7.2$ mbar & $-0.9$ mbar \\ 
\hline
haze optical depth scale factor & HScale & 0.3$\tau_{Jup}$ & $+0.5\tau_{Jup}$ & $-0.3\tau_{Jup}$ \\
\hline
haze profile max density pressure & HSCale & \multicolumn{3}{c|}{no constraint} \\
\hline
haze vertical extent scale factor & HThick& $<10^{6}$ & \multicolumn{2}{c|}{near homogeneous profile} \\
\hline
isothermal temperature profile & T & 480 K & $+145$ K & $-105$ K \\
\hline
HCN mixing ratio & HCN & 0.5 $\%$ & $+0.7$ $\%$ & $-0.4$ $\%$ \\
\hline
CH$_{4}$ mixing ratio & CH4 & 0.5 $\%$ & $+1.1$ $\%$ & $-0.4$ $\%$ \\
\hline
C$_{2}$H$_{2}$ mixing ratio & C2H2 & $<<$ 700 ppm & \multicolumn{2}{c|}{no detection} \\
\hline
CO$_{2}$ mixing ratio & CO2 & $<<$ 0.5 $\%$ & \multicolumn{2}{c|}{no detection} \\
\hline
\end{tabular}
\caption{Atmospheric model constraints estimated from the transmission spectrum by spectral retrieval. The parameter $\tau_{Jup}$ is the opacity of the haze layer in Jupiter (see text for additional discussion). The labels match those used in Figure ~\ref{fig:correlation}.}
\label{tab:retrieval_values}
\end{table*}

\subsection{Previous Observations}

\begin{figure*}[htp!]
\centering
\includegraphics[width=2\columnwidth]{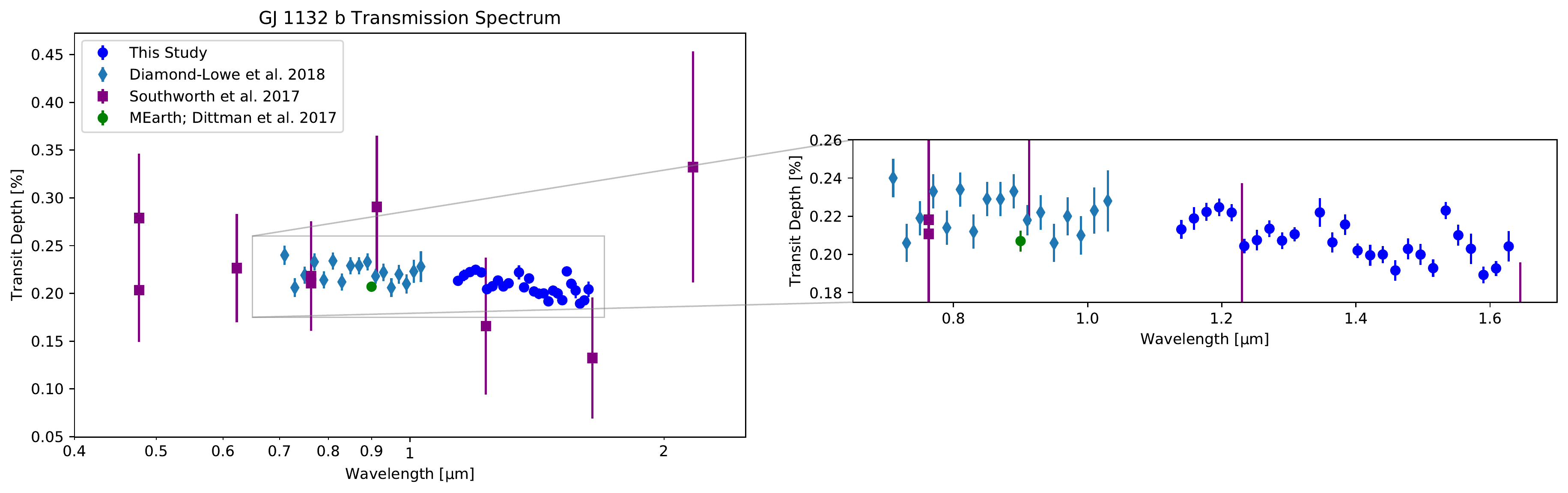}
\caption{The Hubble WFC3-G141 observations reported in this paper are shown with ground-based observations for context. The WFC3 and Southworth observations have been shifted vertically for alignment with the MEarth discovery measurement. The WFC3 data provide better spectral measurement precision than previous ground-based observations at these wavelengths.}
\label{fig:ground}
\end{figure*}

Previous ground-based measurements of GJ 1132 b have reported indications of a H$_{2}$O rich atmosphere \citep{southworth2017} using near-infrared photometry, and a flat spectrum at visible wavelengths \citep{diamond-lowe2018}, the latter being consistent with the presence of aerosol haze or clouds. Our results, which detect an aerosol haze, are consistent with the visible measurements, and our results are also consistent with the ground-based near-infrared photometry finding that the planet has a detectable atmosphere. However, our measurements are inconsistent with aspects of the previous interpretations of ground-based measurements, namely that the atmosphere is H$_{2}$O rich \citep{southworth2017} and H$_{2}$ depleted \citep{diamond-lowe2018}. We note that ground-based measurements face significant challenges \citep{redfield2008,swain2010,pearson2019} in correcting for telluric effects and conclude that the improved measurement precision offered by the multi-epoch Hubble observations explains the difference between our findings and the  previous near-infrared results.

\section{Discussion}

The detection of $\sim$ 250 ppm of spectral modulation in the WFC3 measurements is an observational confirmation that GJ 1132 b hosts a detectable atmosphere. As discussed in the previous section, the nature of the atmosphere can be inferred by retrieval. We turn now to consider the larger context of why GJ 1132 b has an atmosphere. In what follows, we consider atmospheric loss and the implications that the transmission spectrum has for magma composition. We also consider the possible role of tidal heating and consider the potential for HCN production by lightning.  From these separate lines of inquiry, a rich picture of GJ 1132 b emerges. However, one of the most important questions is whether the atmosphere detected is the primary atmosphere, the atmosphere established at the time of formation, or a secondary atmosphere. Thus we begin this section by considering atmospheric loss.

\subsection{Atmospheric Loss}

\begin{figure}[htp!]
\centering
    \subfigure[]{\includegraphics[width=0.47\textwidth]{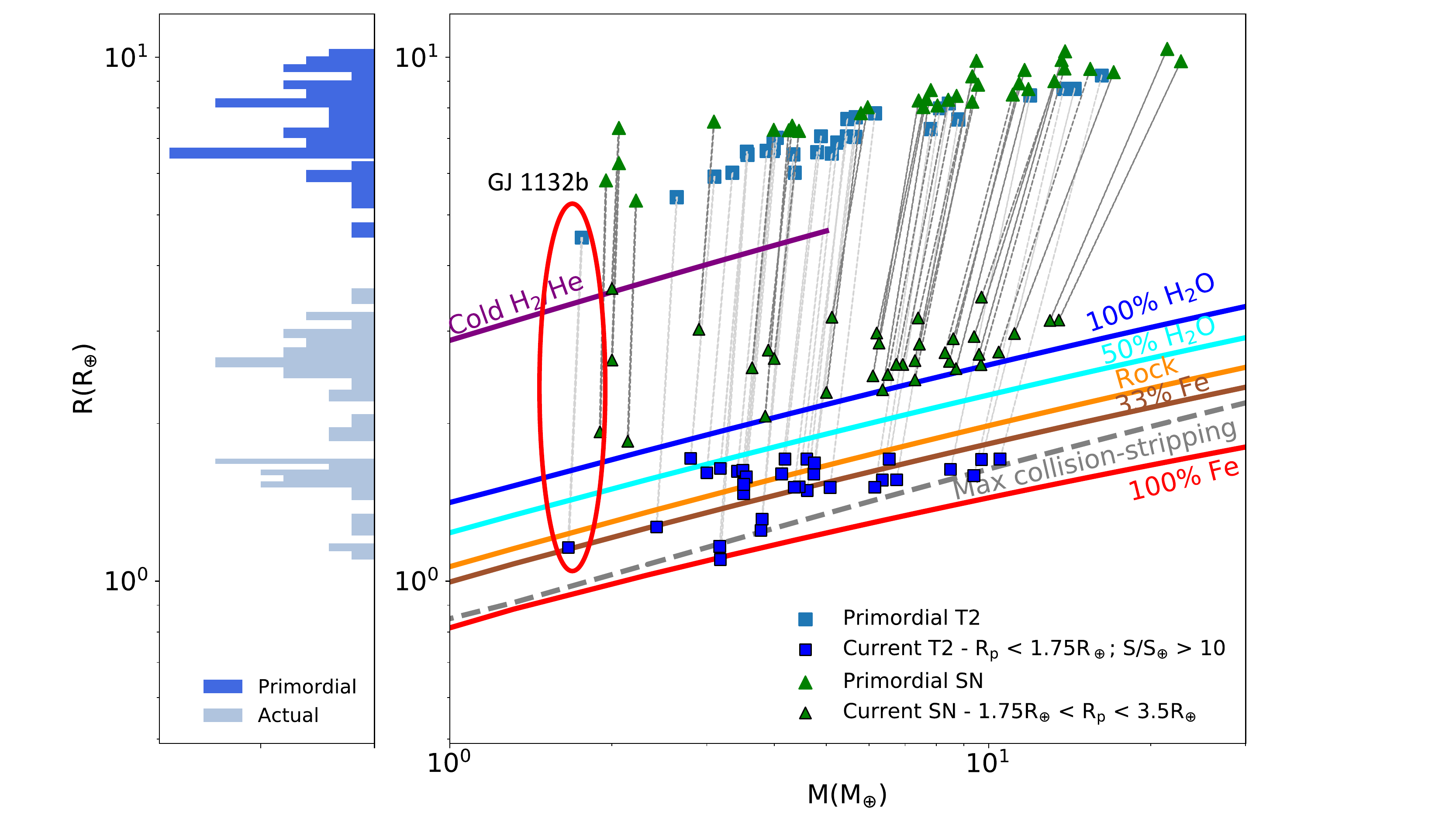}}
    \subfigure[]{\includegraphics[width=0.45\textwidth]{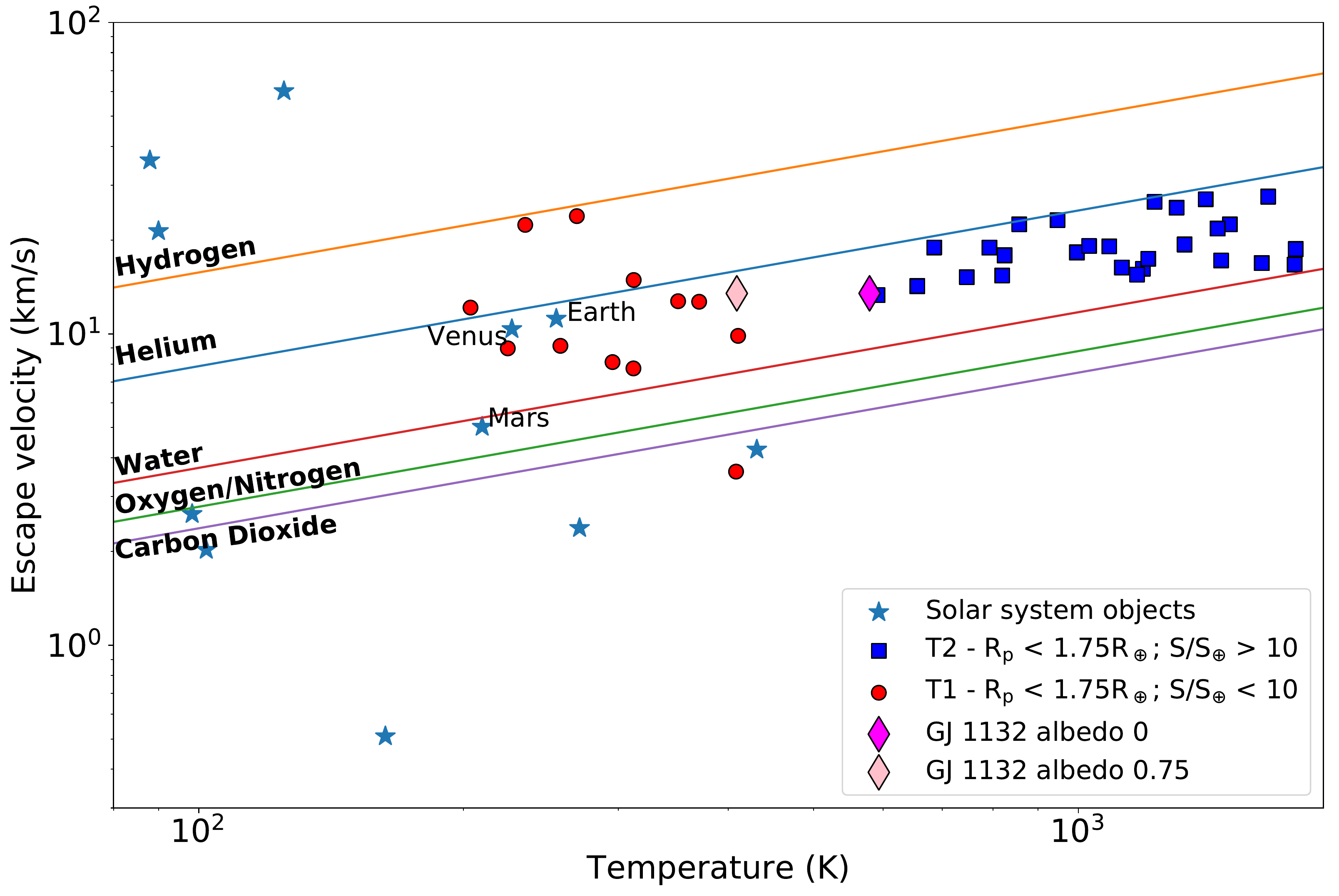}}
\caption{{\bf Top:} GJ 1132 b is a member of a class of strongly insolated terrestrial planets that have likely lost a H/He envelope during the first $\sim100$ Myr, causing significant evolution in the mass-radius diagram. These planets, shown with blue squares, are thought to be the bare cores of what were originally sub-Neptunes. For context, the sub-Neptune planets that have radii R$< $3.5 R$_{\oplus}$ are shown as green triangles. Although the strongly insolated terrestrial planets experience complete envelope loss, the sub-Neptunes experience only partial envelope loss and remain gas-giant-type planets. Updated to highlight GJ 1132 b, this figure originally appeared in \cite{estrela2020} {\bf Bottom:} GJ 1132 b is similar to Earth and Venus and also other terrestrial exoplanets with both low insolation (red) and high insolation (blue) in terms of typical escape velocity. While thermal escape can remove H from the atmosphere of GJ 1132 b, the loss rate is relatively low and H resupply through mantle outgassing could sustain an atmosphere.}
\label{fig:atm_escape}
\end{figure}

Strongly irradiated sub-Neptune mass planets can loose their H/He envelops through hydrodynamic escape \citep{owen2013,fulton2017,owen2017}. Using an energy-limited radiation escape model \citep{murray-clay2009} applicable to this planet, we estimate that GJ 1132 b could have lost a primordial H/He envelope of up to ~2.5$\%$ of the planet mass during the first 100 Myrs of planet life \citep{estrela2020}, corresponding to significant evolution in the mass-radius diagram (see Figure ~\ref{fig:atm_escape}).  In this regard, GJ 1132 b is in family with other strongly irradiated super-Earth planets that are thought to have lost their original H/He evelopes \citep{owen2013,owen2017,fulton2017}.  

Ongoing thermal escape is also a potentially significant loss mechanism we need to consider. The thermal escape of hydrogen occurs when it reaches the escape velocity at the exobase of the planet. The temperature in the exobase will be defined by the XEUV flux of the host star, which varies with the stellar cycles and also in the radiative cooling agents present in the atmosphere of the planet. In the case of exoplanets, we don’t have information yet about the temperature at the exobase. Therefore, here we assumed the equilibrium temperature of the planet providing a lower limit estimation of the thermal atmospheric escape as the exobase temperature can be much higher than the equilibrium temperature due to XUV absorption. Following \cite{konatham2020} we set the thermal escape velocity of the gas species to be one tenth the escape velocity of the planet. Our results are illustrated in Figure ~\ref{fig:atm_escape}, and show that, like Earth and Venus, GJ 1132b (and numerous other small exoplanets) lie below the H and He velocity lines. This means that these terrestrial planets do not have sufficient gravity to retain H/He, and implies that the current loss of H/He of GJ 1132b due to thermal escape is about the same as the Earth or higher, depending on the temperature at the exobase. 

In making these estimates for envelope and present day thermal escape, we are invoking the assumption that the orbit of the planet has not changed substantially since the time of formation. If the orbit of GJ 1132 b has been relatively stable, then it is likely that it lost its primordial H/He envelope. However, if GJ 1132 b has undergone significant orbital migration, especially if that orbital migration occurred relatively recently, then it is possible that the atmosphere detectable today is a remaining component of the primordial atmosphere. Recognizing that orbital migration is a possibility, we proceed to interpret the system under the assumption that the planet was formed and has remained in approximately the current orbital configuration.
  
If the atmosphere of GJ 1132 b contains a significant fraction of H$_{2}$, the position of GJ 1132 b in the mass-radius diagram (Figure ~\ref{fig:atm_escape}) implies that the total amount of atmospheric H$_{2}$ is less than 0.01 \% of the planet mass \citep{lopez2014}. Several hundred times this amount of the H/He primordial envelope would have been lost during the first 100 Myr of the planet’s existence, suggesting any current epoch atmospheric H$_{2}$ must have been resupplied. 

The current atmospheric loss rate from GJ 1132 b has been the subject of Hubble Lyman-$\alpha$ observations that show the H mass loss from GJ 1132 b is below the detectable level of the existing measurements \citep{waalkes2019}. However, the Lyman-$\alpha$ observations do provide input for models and GJ 1132 b's current atmospheric loss rate is estimated to be 3$\times$10$^{7}$ g s$^{-1}$ \citep{waalkes2019}, implying that some resupply may be needed. 

Although the rate of H thermal escape from GJ 1132 b may be modest, the planet is relatively close to the parent star and atmospheric stripping by stellar wind could be substantial. We estimate the wind velocity \citep{parker1960}  using the relation:
\begin{align}
u^{2} - \frac{2kT}{m} - \frac{2kT}{m} \ln \left( \frac{m u^{2}}{2kT} \right) = 8 \frac{kT}{m} \ln \left( \frac{r}{r_{c}} \right) \\ \nonumber
 + 2 G M_{Solar} \left( \frac{1}{r} - \frac{1}{r_{c}} \right)
\end{align}
Where $u$ is the plasma (wind) velocity, $r$ is the distance form the star, $k$ is Boltzmann's constant, $T$ is the plasma temperature, $m$ is the mass of the proton, $r_{c} = G M_{s} m /(4 k T)$ is the critical radius, $G$ is the gravitational constant, and $M_{s}$ is the mass of the star. To model the wind of a M4.5V star with 0.181 solar mass, we use the results of a hydro-dynamic simulation performed for TRAPPIST-1 \citep{dong2018}), an M8V (0.08 solar mass) star, and adjust the coronal temperature \citep{giampapa1996} to 2.3 MK to represent GJ 1132. The mass loss due to stellar wind for a planet without a magnetosphere is: 
\begin{equation}
\dot{M_{a}} \approx 2 \pi R_{p}^{2} \alpha \rho_{w} \nu_{w}
\end{equation}
where $R_{p}$ is the planet radius, $\rho_{w}$ and $\nu_{w}$ are the density and velocity of the stellar wind at the planet location, and $\alpha$ is the entrainment efficiency, which is determined through fitting the mixing-layer model to laboratory experiments, the value of which is between 0.01 and 0.3. For this analysis, we adopted the maximum value of 0.3. If the host star was of solar type, the atmospheric mass loss of GJ 1132b, would be: 2.8x10$^{9}$ g s$^{-1}$. Assuming a wind with constant parameters (density and velocity) throughout the 5 Gyr of the system, this amounts to a total loss of 0.068 M$_{\oplus}$. However, if the central star was a M8 dwarf, like TRAPPIST-1, then the atmospheric mass loss would be 2.7x10$^{6}$ g s$^{-1}$ or 7.2x10$^{-5}$ M$_{\oplus}$ in the lifetime of 5 Gyr. 

To estimate the planetary mass loss due to a M4.5 star, it is necessary to scale the mass loss of the stellar wind by multiplying the wind density by a scaling factor \citep{wood2002}. This factor is the most uncertain assumption and was estimated by a linear interpolation between stellar mass and wind mass loss for the TRAPPIST-1 (M$_{s}$=0.08 M$_{\odot}$, dM$_{\rm w}$/dt=0.001) and the Sun (M$_{\odot}$ and dM$_{\rm w}$/dt=1), yielding a dM$_{\rm w}$/dt$\sim$0.01 wind mass loss for a 0.181 M$_{\odot}$ star. The wind density of TRAPPIST-1 was multiplied by 100 to simulate the wind density of GJ 1132, whereas the wind velocity was calculated using the Parker solution for a 1.7 MK corona and mass of 0.181 M$_{\odot}$. For these conditions, the atmospheric mass loss calculated was 2.2$\times$10$^{8}$ g s$^{-1}$ or 0.0058 M$_{\oplus}$ in the lifetime of 5 Gyr, assuming a constant stellar wind for the entire period. We estimate the current atmospheric stripping by the stellar wind from the $\sim$5 Gyr M4.5 dwarf \citep{dong2018,wood2002} to be $\sim$10 times more efficient than thermal escape unless shielded by a magnetic field.  

Evidence that the atmosphere of GJ 1132 b is a secondary atmosphere is provided by both H$_{2}$, based on primordial envelope loss, the need for ongoing H$_{2}$ replenishment to counter thermal escape, and the detected mixing ratio of HCN. The large HCN mixing ratio ($\sim$0.4\%) requires atmospheric N to be enhanced, potentially by outgassing, above the primordial values, supporting the interpretation that the atmosphere of GJ 1132 b has undergone significant evolution from primordial conditions.  

\subsection{Mantle Outgassing}

Seeking to explore the possible volcanic origin of at least some components of the atmosphere of GJ 1132 b, we used a photo/thermochemical approach to model magma outgassing and atmospheric composition for a wide range of H/He/C/N/O. To constrain the magma conditions, we applied FastChem \citep{stock2018} to solve the equilibrium chemistry over a range of temperature (480 - 2500 K) and pressures (1 bar, 10 bar, 100 bar), and over a wide range of H/He/C/N/O compositions.  

\begin{figure}[htp!]
\centering
\includegraphics[angle=-90,width=0.95\columnwidth]{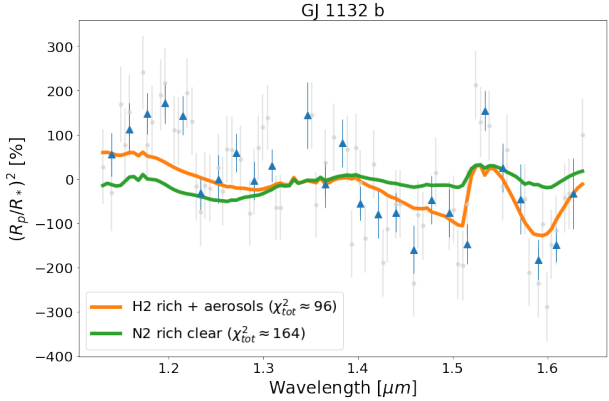}
\caption{A forward model of an H$_{2}$ rich atmosphere based on geochemical modeling inputs is approximately consistent with the observed spectrum. This confirms that the observed spectrum is plausible based on the composition assumptions and the effect of thermo and photochemistry. Forward modeling also shows that a geochemical model solution invoking an N$_{2}$ rich atmosphere produces a relatively flat spectrum, due to the high mean molecular weight, that is inconsistent with the observations.}
\label{fig:H2N2}
\end{figure}

Three classes of solutions were identified that yield an atmospheric composition consistent with the observations, one class for N$_{2}$-dominated magma, a second class for H$_{2}$-dominated magma, and a third class for He-dominated magma. We dismissed the third class because it is unlikely that a planet of GJ 1132 b’s mass would be Helium dominated \citep{hu2015}, and we do not attempt to model incorporating He-dominated gas within its mantle. The remaining atmospheric classes were separately applied as a surface boundary condition, assuming an adiabatic atmosphere for pressures greater than 10$^{-2}$ bars, a  isothermal atmosphere with T = 480 K for pressures less than 10$^{-2}$ bars, $p_{0}$ = 1 bar, $K_{zz}$ = 10$^{7}$ cm$^{2}$  s$^{-1}$, and irradiation by GJ 436 as a proxy for GJ 1132 b. The composition inferred is indicative of the surface pressure of the planet \citep{gaillard2014}, the temperature of the magma, and the redox state of the mantle/xenolith, described by its oxygen fugacity. The N$_{2}$-dominated magma class of models produces an atmosphere that is $\sim70$\% N$_{2}$ with mean molecular weight of $\sim21$. To determine if atmospheres of this kind are consistent with the WFC3 observations, we explored the N$_{2}$ dominated atmosphere hypothesis. As shown in Figure ~\ref{fig:H2N2}, an N$_{2}$ rich atmosphere provides a relatively poor fit to the observations when compared with the H$_{2}$ rich atmosphere produced by H$_{2}$-dominated magma. We therefore concentrate on the H$_{2}$ rich scenario in the subsequent discussion.

We take the best fit results of FastChem and apply them to a 1D photochemistry-diffusion code ARGO \citep{rimmer2016} to solve the atmospheric transport equation for the steady-state vertical composition of GJ 1132 b. ARGO is a Lagrangian photochemistry/diffusion code that follows a single parcel as it moves vertically through the atmosphere. The temperature, pressure, and actinic ultraviolet flux are set at each height in the atmosphere. In this reference frame, bulk diffusion terms are accounted for by the time dependence of the chemical production, $P_{i}$ ($\rm cm^{3} s^{-1}$), and loss, Li ($\rm s^{-1}$), and so below the homopause, the chemical equation being solved is effectively:
\begin{equation}
\frac{\partial n_{i}}{\partial t} = P_{i} \left[t(z,\nu_{z})\right] - L_{i} \left[t(z,\nu_{z})\right] n_{i}
\end{equation}
where $n_{i}$ (cm$^{-3}$) is the number density of species $i$, $t($s$)$ is time, $z$ [cm] is atmospheric height, and $v_{z}=K_{zz} /H_{0}$ (cm/s) is the effective vertical velocity due to Eddy diffusion, from the Eddy diffusion coefficient $K_{zz}$ (cm$^{2}$ s$ ^{-1}$). The model is run until the concentration of every species with $n_{i}>1$ cm$^{-3}$  does not change by more than 1\% between two global iterations. The chemical network we use is STAND2018 \citep{rimmer2016,rimmer2019}.

The model results predict observable quantities of HCN, CH$_{4}$ and possibly C$_{2}$H$_{2}$ above 1 mbar, as well as significant amounts of HC$_{3}$N. In the modeling results, the haze is the sum of C$_{x}$H$_{y}$ ($x \geq$ 3) and C$_{x}$H$_{y}$N$_{z}$ terminating species – for details see \citep{rimmer2016} – for which there are other homologation reactions that are not included in the network, and that are known precursors to haze particles. These are C$_{3}$H$_{4}$, C$_{4}$H$_{4}$, CH$_{2}$CN, C$_{2}$H$_{3}$N, and C$_{3}$H$_{3}$N. Although we do not model haze formation, we track the mixing ratio of 'haze', which is the sum of the mixing ratios of the aforementioned species. It provides a qualitative indicator of where the haze probably forms, but not any of its other properties.

We created forward models for the atmospheric transmission spectra based on the N$_{2}$-dominated magma and  H$_{2}$-dominated magma classes of solutions. The  H$_{2}$-dominated magma classes of solutions is favored (see Figure ~\ref{fig:H2N2}) and produces an atmospheric composition that is similar to the atmospheric composition inferred from the free retrieval of the Hubble WFC3 G141 data. The N$_{2}$-dominated case leads to a higher mean molecular weight atmosphere and a relatively flat transmission spectrum that is inconsistent with the measurement.

\begin{figure}[htp!]
\centering
\includegraphics[width=1\columnwidth]{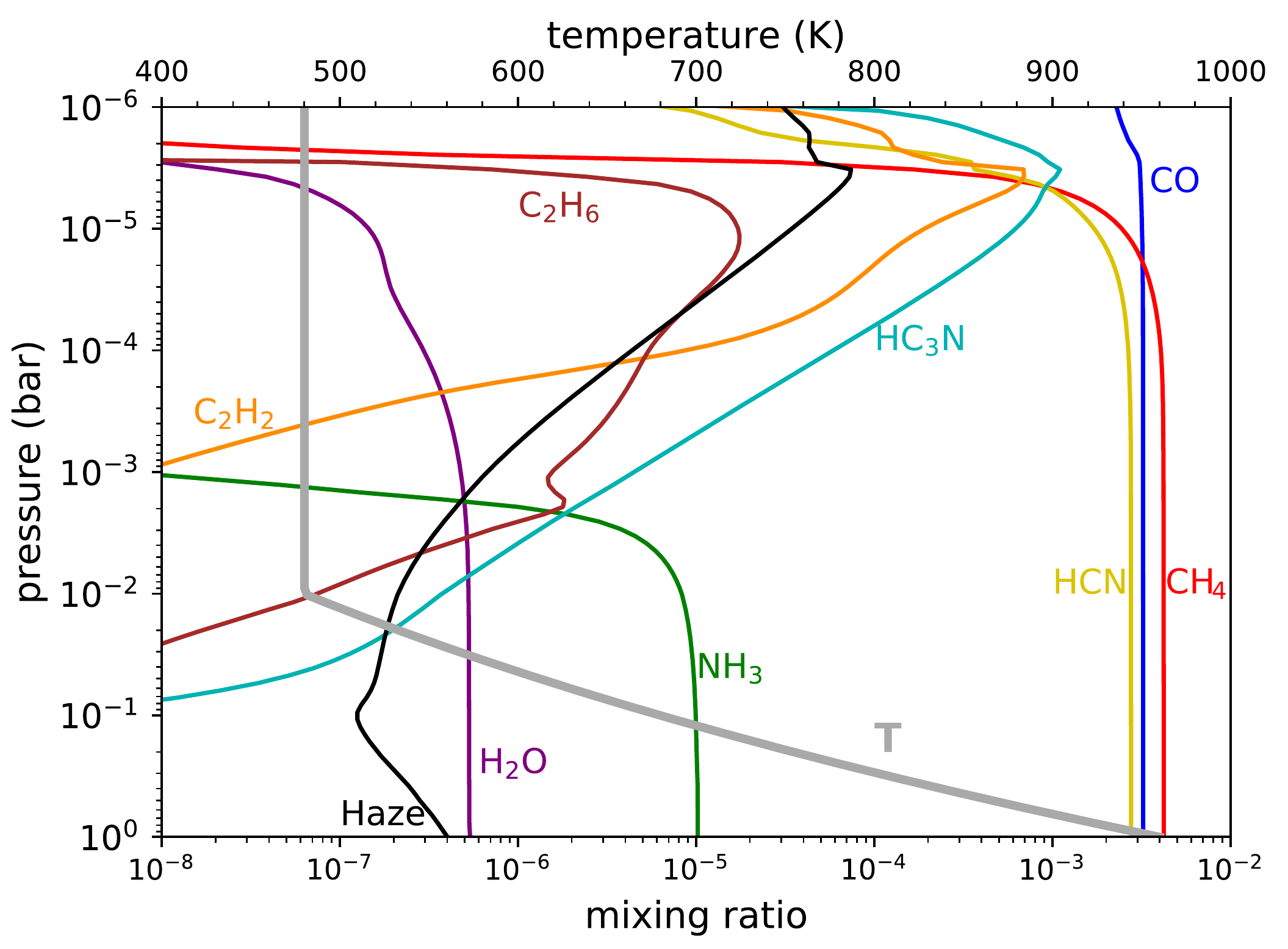}
\caption{Atmospheric profiles of species originally produced by geochemical outgassing and then modified by thermal equilibrium chemistry, according to the pressure temperature profile (shown in grey), and photochemistry. In the case of GJ 1132 b, the inference from observing HCN and CH$_{4}$ is that these are volcanic gases produced by active ultrareduced volcanism that has converted the atmosphere to volcanically-derived composition. The haze precursor species are plotted in black. The model parameters plotted here correspond to the values given in Table ~\ref{tab:outgass_model}.}
\label{fig:profiles}
\end{figure}

To summarize, we explored a wide range of H/He/C/N/O values and found two families of solutions, H$_{2}$ and N$_{2}$ dominated respectively, that produced outgassing results leading to the production of CH$_{4}$ and HCN  inferred by the retrieval. In both cases, C/O is enhanced with respect to solar values, consistent with the inferences above. However, as already discussed, an N$_{2}$ dominated atmosphere results in a transmission spectrum that is inconsistent with the observations and we exclude the N$_{2}$ dominated case from further consideration. The major products from the H$_{2}$ dominated magma outgassing model category are shown in Table ~\ref{tab:outgass_model}, and the vertical mixing ratio for these products is shown in Figure ~\ref{fig:profiles}. The magma outgassing products include $\sim$8\% N$_{2}$, which is compatible with the atmospheric mean molecular weight constraints based on the near-infrared spectrum. In addition to producing results consistent with the observed amounts of HCN, CH$_{4}$, and the implied H$_{2}$ content, modeling the magma outgassing composition also produces atmospheric haze precursor species C$_{x}$H$_{y}$ and C$_{x}$H$_{y}$N$_{z}$; these products have been identified as haze precursors on Titan \citep{lavvas2008} and would be consistent with the interpretation of the overall spectral slope as due to aerosol scattering. We confirmed the need for an aerosol contribution to the H$_{2}$, HCN, CH$_{4}$ atmospheric opacity by testing an aerosol free atmosphere model and found the disagreement between the aerosol free atmosphere model and the data becomes significant in the red portion of the spectrum (see top portion of Figure ~\ref{fig:water}). By generating atmospheric transmission models based, we show the geochemical model prescription, including the results for C$_{2}$H$_{2}$ and NH$_{3}$, can be made consistent with the observations  (see Figure ~\ref{fig:H2N2}) by the inclusion of a combination of haze and cloud with properties similar to what was found in the free retrieval based on the observed spectrum. 

\begin{table}[htp!]
\centering
\begin{tabular}{|c|c|}
\hline
\textbf{Parameter} & \textbf{Value} \\ 
\hline 
surface pressure & 1 bar \\
\hline
outgassing temperature & 1670 K \\
\hline
atmospheric mean molecular weight & 4.54 \\
\hline
C/O ratio & 3.16 \\
\hline
H$_{2}$ mixing ratio & 90 $\%$ \\
\hline
He mixing ratio & 0.1 $\%$ \\
\hline
HCN mixing ratio & 0.3 $\%$ \\
\hline
C$_{2}$H$_{2}$ mixing ratio & 0.06 $\%$ \\
\hline
CH$_{4}$ mixing ratio & 0.3 $\%$ \\
\hline
CO mixing ratio & 0.3 $\%$ \\
\hline
CO$_{2}$ mixing ratio & 0.58 ppb  \\
\hline
H$_{2}$O mixing ratio & 530 ppb \\
\hline
N$_{2}$ mixing ratio & 8.9 $\%$ \\
\hline
NH$_{3}$ mixing ratio & 10.2 ppm \\
\hline
He mixing ratio & 0.1 $\%$ \\
\hline
log$f$O$_{2}$ & IW-11 \\
\hline 
\end{tabular}
\caption{Outgassing model selected parameters for the H$_{2}$ dominated atmosphere case. The model parameter values represent the atmosphere at the outgassing location; thermochemistry and photochemistry modify the vertical mixing ratio of atmospheric constituents as a function of temperature and pressure.}
\label{tab:outgass_model}
\end{table}

\subsection{Testing the Ultrareduced Volcanism Hypothesis}

Although the scenario of ultrareduced volcanism developed above is capable of explaining both the atmospheric origin and composition inferred from the Hubble WFC3 observations, the observational evidence is based on measurements with limited spectral coverage. Confirming the ultrareduced volcanism hypothesis is an important area for further study because of the possibility that it could be more widely associated with reestablishing atmospheres on other powerfully irradiated super-Earths.

\begin{figure}
    \centering
    \subfigure[]{\includegraphics[width=0.45\textwidth]{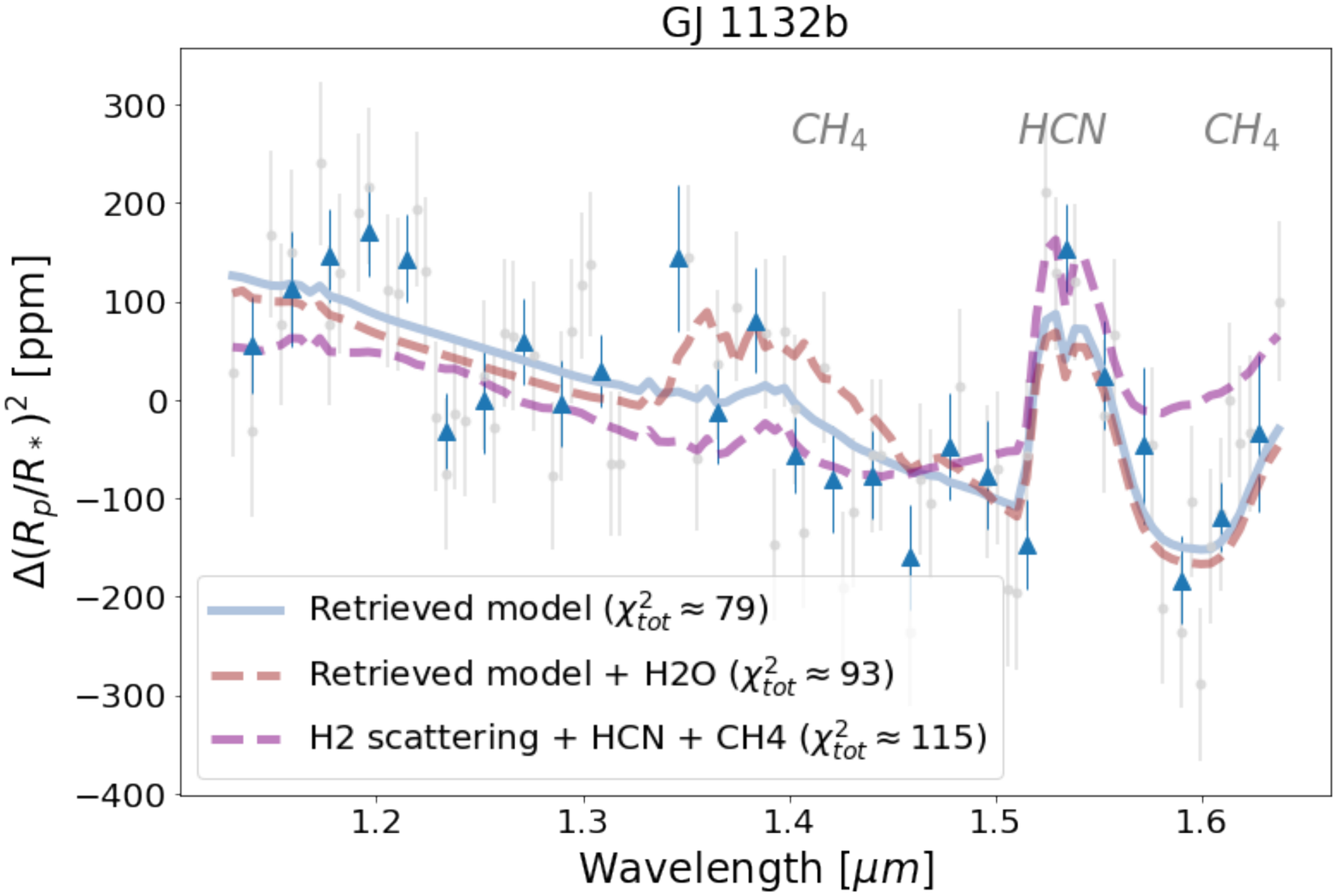}}
    \subfigure[]{\includegraphics[width=0.45\textwidth]{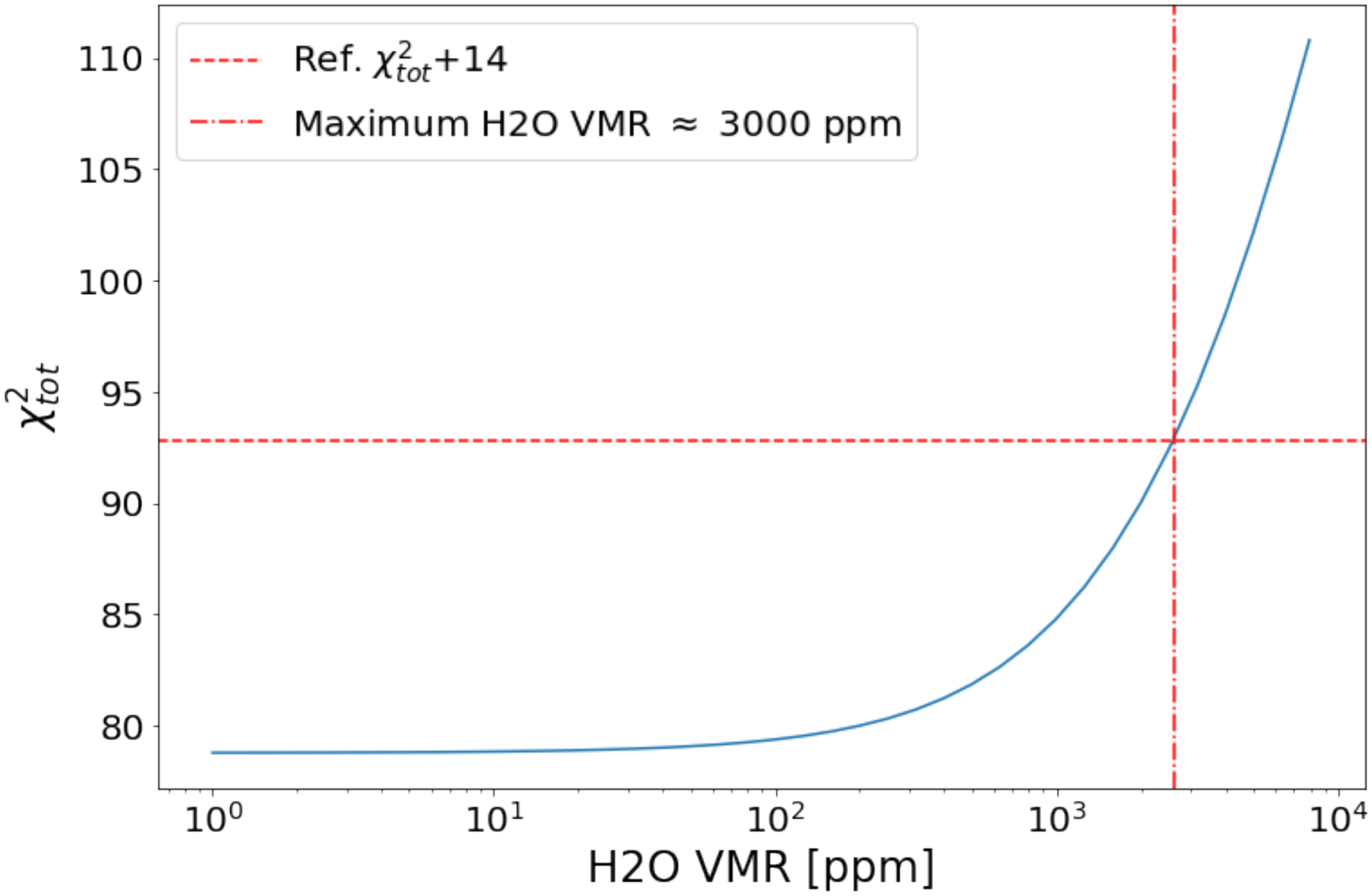}}
    \caption{{\bf Top:} A clear atmosphere model is not favored by the data. Although  the data are most consistent with an absence of water on a $\chi^{2}$ basis, some amount of H$_{2}$O could be hidden by the atmospheric haze in the retrieved model.  {\bf Bottom:} The results of considering a range of H$_{2}$O vertical mixing ratio values indicates that perhaps as much as 3000 ppm of H$_{2}$O could be consistent with the observed spectrum. }
    \label{fig:water}
\end{figure}

For the ultrareduced volcanism model proposed for GJ 1132 b, the most stringent and most easily executable observational test is to search for H$_{2}$O in GJ 1132 b's atmosphere. The ultrareduced volcanism model makes explicit predictions about the O fugacity of the magma, the C/O ratio, and the atmospheric mixing ratio of H$_{2}$O. The predicted 530 ppb H$_{2}$O mixing ratio is an extremely low value and any detection of substantial larger amounts of H$_{2}$O in the atmosphere of GJ 1132 b would challenge the ultrareduced volcanism hypothesis.  We explored the range of H$_{2}$O mixing ratios consistent with the WFC3 observations and  Figure ~\ref{fig:water} shows an atmospheric model incorporating 2600 ppm of water and the results of varying the vertical mixing ratio for water through a wide range of values. This analysis indicates that H$_{2}$O vertical mixing ratios of less than $\sim$ 3000 ppm are potentially consistent with the observations. Thus, the Hubble WFC3 data, given the presence of the aerosol, are not sufficient to provide useful constraints on the H$_{2}$O mixing ratio. Although not explicitly detected, H$_{2}$O could be consistent with the measured spectrum if present alongside CH$_{4}$ (see Figure ~\ref{fig:water}) and the possibility of significant atmospheric water has been studied \citep{schaefer2016}. Determining the atmospheric mixing ratio of H$_{2}$O is an important area for further study, and observations with the James Webb Space Telescope would be able to decisively establish the amount of H$_{2}$O present in the atmosphere of GJ 1132 b.

\subsection{Lightning Generated HCN}
As an alternative to the volcanism hypothesis, we considered the possibility of HCN production due to lightning discharges.  HCN is known to be produced by lightning in reduced atmospheres. Assuming that CH$_{4}$ degassing is balanced by diffusion-limited escape, the degassing of CH$_{4}$ to maintain observed concentrations is $\Phi$(CH$_{4}$) $\sim$10$^{13}$ cm$^{-2}$ s$^{-1}$. Assuming half of the degassed CH$_{4}$ is converted into HCN, and given the conversion rate of N$_{2}$ and CH$_{4}$ into HCN from of P $\approx$ 10$^{17}$ molecules J$^{-1}$ \citep{chameides1981}, and typical Earth flash energies of E = 5 × 10$^{9}$ J \citep{romps2014}, this would imply a lightning flash rate of: 
\begin{equation}
F={\rm \frac{\Phi(CH_4 )}{2EP}} = \frac{5 \times 10^{12} {\rm cm^{-2} s^{-1}}} {(5 \times 10^{9} {\rm J}) \times (10^{17} {\rm molecules/J}) )}
\end{equation}
The implied flash rate is 0.36 km$^{-2}$ h$^{-1}$, several thousand times greater than the global average lightning flash rate on Earth \citep{hodosan2016} and, on this basis, we conclude that volcanic outgassing origin for the HCN is more likely.

\subsection{Tidal Dissipation}

The origin of the secondary atmosphere on GJ 1132 b may have been influenced by tidal heating.  The planet has an eccentricity \citep{bonfils2018} of $<$ 0.22, which would normally be rapidly circularized but, in this case, could be maintained by the 11/2 resonance with GJ 1132 c. A nonzero eccentricity potentially leads to significant tidal heating. For example, for an interior model similar to Earth (using the same same value of the imaginary part of the tidal Love number)  and an eccentricity of 0.01, the tidal heat flux is 80 W m$^{-2}$, which is about 10$^{3}$ times larger than the Earth’s internal heat flux. Although large compared to the Earth, an interior heat flux of 80 W m$^{-2}$ is less than the stellar flux received by GJ 1132 b and, in this scenario, the surface temperature would still be controlled by the atmosphere and stellar flux. 

Assuming an Earth-like thermal conductivity of 3 W m$^{-1}$ K$^{-1}$, an interior heat flux of 80 W m$^{-2}$ results in an outer solid layer of a few 10s of meters overlying a magma ocean. The presence of a magma ocean would act to increase the value of the imaginary part of the tidal Love number to
make it closer to the value for Io, increasing the tidal heating and reducing the thickness of the outer solid layer. Volcanic activity resembling that at Io is therefore predicted. If there is a resonance
with GJ 1132 c, the tidal dissipation in GJ 1132 b is much larger than the heating due to radioactive decay, which is the main heat source for planets like Earth, Mars, and Venus. We conclude that even with an eccentricity of 0.01 (much lower than the observational maximum value of 0.22), GJ1132 b should be volcanically active. If H$_{2}$ was dissolved in the magma ocean during formation, it could be outgassed due to tidal-dissipation-induced volcanic activity and is a plausible origin for the H$_{2}$-dominated secondary atmosphere.  

\subsection{Discussion Summary}

Although GJ 1132 b has a similar size and density to Earth, it has had a very different evolutionary trajectory. If GJ 1132 b is representative of other strongly irradiated super-Earths, we can infer that it started life with a hydrogen envelope, and these circumstances caused significant amounts of H$_{2}$ to rapidly ingass into the mantle \citep{chachan2018,kite2019}. This formed a reservoir of H$_{2}$ that is now degassing at a rate sufficient to maintain a moderately reducing (H$_{2}$, N$_{2}$, CH$_{4}$, and probably CO) atmosphere. Degassing via high-temperature volcanism, which we model as occurring at 1 bar pressure and 1670 K magma, could result in quantities of HCN, CH$_{4}$, NH$_{3}$, and C$_{2}$H$_{2}$ that are consistent with the observations. Because we do not know the surface pressure or where the atmosphere becomes adiabatic, it is difficult to constrain the surface temperature and we cannot distinguish between surface or subsurface degassing. 

The photochemically-derived vertical mixing ratio profile for NH$_{3}$ (photodissociated above $\sim$ 1 mbar), the presence of a cloud or photochemical haze at $\sim$1 mbar pressure, and the inability to probe pressures much greater than $\sim$ 1 mbar in transit observations at these wavelengths all act to minimize the signature of any NH$_{3}$ in the measured spectrum. Our interpretation is that GJ 1132 b is in the volcanic degassed phase, but because of its very different geological history from Earth, GJ 1132 b’s volcanic degassed phase is much more chemically reduced. Rocks derived from the upper mantle under Mount Carmel, Israel, demonstrate that the low oxygen fugacity inferred by our models (IW-10 $<$ fO$_{2}$ $<$ IW-12) is within the values of fO$_{2}$ $<$ IW-10 within certain local regions of the Earth’s upper mantle \citep{griffin2016}. Since local H$_{2}$ ingassing provides the best explanation for the low oxygen fugacity values of the Mount Carmel samples \citep{griffin2016}, global H$_{2}$ ingassing would plausibly provide similar oxygen fugacities needed for volcanism to produce the observed quantities of HCN and CH$_{4}$. In the case of Earth, hydrogen was lost to space and the reducing power of iron was sequestered in the core. We propose that GJ 1132 b lost some fraction of primordial hydrogen to the upper mantle and is now slowly degassing that hydrogen and losing it to space.

We conclude some H regeneration mechanism is needed because atmospheric loss modeling of GJ 1132 b strongly suggests the primordial H/He envelope was lost. \cite{kite2020} predicted a secondary atmosphere for GJ 1132 b, and we observe an atmosphere that likely arose through outgassing. However, the mean molecular weight of the atmosphere we observe differs from the \cite{kite2020} prediction and implies the mantle has remained melted. We propose a modified version of the \cite{kite2020} scenario for GJ 1132 b in which the mantle does not crystallize (due to tidal heating) but supports a relatively thin solid crust potentially 10s of meters in thickness. The solid crust slows the loss of H from the magma reservoir so that at the $\sim 5$ Gyr planet age, there is still some H to degas, and a solid crust is consistent with the dayside planet emission being undetected by Spitzer secondary eclipse measurements. 

Our combination of geochemical and photochemical atmospheric modeling demonstrates a way to connect the degassed volatile content to atmospheric chemistry. Assuming the volcanism scenario is correct, the combined geochemical and photochemical modeling uses the observed transit spectrum to derive some constraints on mantle properties, namely, the oxygen fugacity (very low), the hydrogen fugacity (very high), and the magma temperature (if the temperature is too low, N$_{2}$ and CH$_{4}$ are produced without any HCN). But our modeling doesn't complete the mantle-atmosphere connection because the model lacks a self-consistent treatment of the mantle chemistry. A area for future investigations is improved modeling of the mantle to explore effects such as the role of dissolved N on oxygen fugacity \citep{libourel2003}.

\section{Conclusions}

Assuming that it has not undergone recent and significant orbital migration, GJ 1132 b is a terrestrial planet that lost its primordial H/He envelope and has since re-established an atmosphere that contains observable signatures of an aerosol, CH$_{4}$, and  HCN, while H$_{2}$ is inferred from spectral modulation and modeling. The present epoch secondary atmosphere, including H$_{2}$, can be explained by mantle outgassing. A plausible scenario is one involving volcanic activity, facilitated by tidal heating, releasing H that was originally dissolved in the primordial magma ocean during the planetary accretion phase. Alternatively, the atmospheric volatiles could have been delivered by bombardment. The sloped spectrum is likely due to Rayleigh scattering by small particles that are present at altitudes corresponding to mbar pressures.   An important question for future investigation is determining the H$_{2}$O mixing ratio and the C/O ratio, which will help refine our understanding of the mechanism that produced the secondary atmosphere on GJ 1132 b. In particular, transit observations with the James Webb Space Telescope, specifically aimed at probing the abundance of CO, CO$_{2}$, and searching for H$_{2}$O bands, can measure atmospheric C/O ratio and provide tight constraints on the outgassing chemistry and magma composition.

Our findings have important implications for future exoplanet observational studies. While terrestrial planets that were stripped of their envelopes might appear, prima facie, as poor choices for transit spectroscopy observations (because the atmosphere has either been removed or converted into a high-mean-molecular-weight atmosphere), our findings suggest that significant H outgassing can make high-insolation terrestrial planets desirable observational targets. Secondary atmospheres, as a class, probe a rich range of possible physical and chemical processes and offer a high degree of potential complexity, including the possibility of bio-signatures on habitable worlds. Numerous high-insolation super-Earth planets are currently known, and this population may play an important role in the future study of secondary atmospheres. UV stripping of H leads to significant enhancement of D/H, as in the case of Venus, raising the possibility of future searches for rotational and ro-vibrational HD lines in the 2.5 to 38 $\mu$m wavelength range, and potentially HDO. An important question raised by this work is, does ingassing of H-rich envelope material during the formation epoch produce significant consequences on mantle oxygen fugacity and profoundly impact the longer-term abundance of oxidized species? Given the currently known potential targets, and the rich variety of chemical and physical processes in play, it is likely that study of terrestrial secondary atmospheres will become a fruitful area of exoplanet research in the near future.

\section*{Acknowledgments}

The authors thank Zach Berta-Thomson for leading the Hubble observing program (PID 14758) which provided the data used in this study.  The authors also thank the anonymous referee for comments that improved the presentation of results reported in this manuscript. This research has made use of the NASA Exoplanet Archive, which is operated by the California Institute of Technology, under contract with the National Aeronautics and Space Administration under the Exoplanet Exploration Program. Mark Swain, Raissa Estrela, and Kyle Pearson acknowledge support for a portion of this effort from NASA ADAP award 907524. Raissa Estrela acknowledges Sao Paulo Research Foundation (FAPESP) for the fellowship \#2018/09984-7. Paul B. Rimmer thanks the Simons Foundation for funding (SCOL awards 599634). This work has been supported in part by the California Institute of Technology Jet Propulsion Laboratory Exoplanet Science Initiative. This research was carried out at the Jet Propulsion Laboratory, California Institute of Technology, under a contract with the National Aeronautics and Space Administration. © 2021.
All rights reserved. 

\software{LDTK \citep{parviainen2015}, PyMC3 \citep{salvatier2016}, ARGO \citep{rimmer2016}, FastChem\citep{stock2018}}
\bibliography{references}

\section*{Appendix}

\begin{table*}[htp!]
\centering
\begin{tabular}{|c|c|c|}
\multicolumn{3}{c}{\textbf{Table of Spectral Values - Part A} }\\ \hline
\textbf{wavelength in $\mu$m} & \textbf{offset in ppm}  & \textbf{1 $\sigma$ in ppm}\\ \hline
1.130 & 27.4 & 84.3 \\
1.149 & 168.2 & 85.2 \\
1.153 & 76.3 & 81.9 \\
1.158 & 150.8 & 83.7 \\
1.172 & 240.3 & 83.1 \\
1.177 & 76.2 & 81.4 \\
1.181 & 128.5 & 81.3 \\
1.191 & 190.3 & 80.5 \\
1.196 & 216.6 & 79.6 \\
1.205 & 110.9 & 77.6 \\
1.210 & 107.3 & 77.7 \\
1.219 & 193.6 & 78.0 \\
1.224 & 130.5 & 76.1 \\
1.228 & -17.1 & 76.3 \\
1.233 & -75.3 & 76.9 \\
1.238 & -14.6 & 75.5 \\
1.242 & -21.6 & 75.7 \\
1.252 & 24.8 & 76.1 \\
1.257 & -27.3 & 77.2 \\
1.261 & 67.4 & 75.4 \\
1.266 & 65.2 & 76.0 \\
1.275 & 44.6 & 75.6 \\
1.285 & -77.4 & 74.9 \\
1.289 & -6.3 & 75.2 \\
1.294 & 70.2 & 73.3 \\
1.299 & 116.4 & 74.4 \\
1.303 & 138.2 & 73.4 \\
1.313 & -65 & 73.0 \\
1.318 & -64.4 & 72.8 \\
1.350 & 144.1 & 74.8 \\
1.355 & -59.1 & 74.0 \\
1.364 & 35.9 & 75.7 \\
1.374 & 94.5 & 75.7 \\
1.388 & 67.1 & 76.1 \\
1.393 & -147.1 & 76.6 \\
1.397 & 70.1 & 77.1 \\
1.402 & -9.4 & 75.4 \\
1.407 & -135.0 & 76.2 \\
1.416 & 33.0 & 77.2 \\
\hline
\hline
\end{tabular}
\caption{Data for Figure ~\ref{fig:spectrum} from 1.130 to 1.416 $\mu$m where offset is respect to the broadband transit depth. Corresponds to grey points in Figure ~\ref{fig:spectrum}}
\label{tab:spectrum_values}
\end{table*}

\begin{table*}[htp!]
\centering
\begin{tabular}{|c|c|c|}
\multicolumn{3}{c}{\textbf{Table of Spectral Values - Part B} }\\ \hline
\textbf{wavelength in $\mu$m} & \textbf{offset in ppm}  & \textbf{1 $\sigma$ in ppm}\\ \hline
1.426 & -189.7 & 76.1 \\
1.430 & -114 & 75.6 \\
1.440 & -56.4 & 77.4 \\
1.444 & -56.3 & 76.9 \\
1.458 & -235.8 & 75.2 \\
1.463 & -80.5 & 76.5 \\
1.468 & -105.8 & 76.1 \\
1.482 & 14.5 & 78.0 \\
1.496 & -83.5 & 77.4 \\
1.501 & -69.3 & 77.1 \\
1.505 & -192.7 & 78.2 \\
1.510 & -195.8 & 78.4 \\
1.515 & -56.2 & 77.1 \\
1.524 & 211.6 & 77 \\
1.529 & 128.3 & 78.1 \\
1.538 & 120 & 79.5 \\
1.552 & -16.2 & 77.9 \\
1.557 & 65.8 & 77.7 \\
1.576 & -46.0 & 79.1 \\
1.580 & -211.2 & 77.0 \\
1.590 & -235.5 & 78.0 \\
1.594 & -102.9 & 77.1 \\
1.599 & -288.8 & 78.1 \\
1.604 & -148.2 & 78.1 \\
1.609 & -116.1 & 77.7 \\
1.613 & 0.0 & 79.1 \\
1.618 & -42.9 & 77.9 \\
1.623 & -33.5 & 79.4 \\
1.637 & 99.9 & 80.7 \\
\hline
\hline
\end{tabular}
\caption{Data for Figure ~\ref{fig:spectrum} for 1.426 to 1.637 $\mu$m where offset is respect to the broadband transit depth. Corresponds to grey points in Figure ~\ref{fig:spectrum}.}
\label{tab:spectrum_values}
\end{table*}

\begin{table*}[htp!]
\centering
\begin{tabular}{|c|c|c|}
\multicolumn{3}{c}{\textbf{Table of Lower Spectral Resolution Spectral Values} }\\ \hline
\textbf{wavelength in $\mu$m} & \textbf{offset in ppm}  & \textbf{1 $\sigma$ in ppm}\\ \hline
1.139 & 55.4 & 49.2 \\
1.158 & 112.7 & 58.5 \\
1.177 & 147.1 & 47.0 \\
1.196 & 171.4 & 45.9 \\
1.214 & 143.5 & 44.9 \\
1.233 & -31.9 & 38.0 \\
1.252 & -0.8 & 54.2 \\
1.271 & 59.1 & 43.6 \\
1.289 & -3.3 & 43.2 \\
1.308 & 30.1 & 36.6 \\
1.346 & 144.1 & 74.8 \\
1.364 & -12.6 & 52.9 \\
1.383 & 80.9 & 53.6 \\
1.402 & -55.5 & 38.2 \\
1.421 & -80 & 54.2 \\
1.440 & -76.1 & 44.4 \\
1.458 & -159.5 & 53.6 \\
1.477 & -47.0 & 54.5 \\
1.496 & -76.3 & 54.6 \\
1.515 & -147.3 & 45.1 \\
1.533 & 154.3 & 45.1 \\
1.552 & 24.9 & 55.0 \\
1.571 & -46.0 & 79.1 \\
1.590 & -182.8 & 44.5 \\
1.609 & -119.7 & 34.9 \\
1.627 & -33.5 & 79.4 \\
\hline
\hline
\end{tabular}
\caption{Data for Figure ~\ref{fig:spectrum} for 1.426 to 1.637 $\mu$m where offset is respect to the broadband transit depth. Corresponds to blue points in Figure ~\ref{fig:spectrum}.}
\label{tab:spectrum_values}
\end{table*}

\end{document}